\documentclass[11pt]{article}
\usepackage{apacite}
\usepackage{natbib}
\usepackage{amsmath}
\usepackage{amsfonts}
\usepackage{amssymb}
\usepackage{graphicx}
\usepackage[lmargin=2cm,rmargin=2.5cm]{geometry}
\usepackage{empheq}
\usepackage{xcolor}
\usepackage{hyperref}
\definecolor{PPgreen}{HTML}{70AD47}
\definecolor{PPred}{HTML}{c00000}
\definecolor{PPblue}{HTML}{4472C4}
\newcommand{\re}[1]{#1}

\title{\LARGE\bf
Bayesian parameter estimation for the SWIFT model of eye-movement control during reading} 
\author{Stefan A.~Seelig$^1$, Maximilian M. Rabe$^1$, Noa Malem-Shinitski$^2$, Sarah Risse$^1$, \\ 
Sebastian Reich$^{2,3}$, \& Ralf Engbert$^{1,3}$ 
\\ $^1$Department of Psychology, $^2$Institute of Mathematics,\\$^3$Research Focus Cognitive Sciences, \\ 
University of Potsdam, Germany}

\begin{document}
\maketitle
\abstract{Process-oriented theories of cognition must be evaluated against time-ordered observations. Here we present a representative example for data assimilation of the SWIFT model, a dynamical model of the control of fixation positions and fixation durations during \re{natural reading of single sentences.} First, we develop and test an approximate likelihood function of the model, which is a combination of a spatial, pseudo-marginal likelihood and a temporal likelihood \re{obtained by probability density approximation.} Second, we \re{implement} a Bayesian approach to parameter inference using an adaptive Markov chain Monte Carlo procedure. Our results indicate that model parameters can be estimated reliably for individual subjects. We conclude that approximative Bayesian inference represents a considerable step forward for \re{computational models} of eye-movement \re{control,} where modeling of individual data on the basis of process-based dynamic models has not been possible \re{so far.}
\\
[2ex]
{{\sl Keywords:} Dynamical models, reading, eye movements, saccades, likelihood function, Bayesian inference, MCMC, interindividual differences}

\section{Introduction}
Dynamical models represent an important theoretical approach to cognitive systems, in particular, if we seek to explain time-ordered behavioral data such as sequences of movements. In dynamical models, sequential dependencies between observations are naturally explained by underlying dynamical principles that unfold over time when the model is simulated numerically \citep{vanGelder1998,Beer2000}. Examples for the dynamical approach can be found in many fields of cognitive research, triggered by early examples from motor control \citep{Haken1985,Erlhagen2002} or decision field theory \citep{Busemeyer1993}.  

Dynamical models generate highly specific predictions on sequential data that include statistical correlations between the subsequent observations over time. As a consequence, parameter inference for dynamical models must be carried out with the fully dynamical framework of {\sl data assimilation} \citep{Law2015,Reich2015}. Here we investigate parameter inference in the SWIFT model of saccade generation during reading~\citep{Engbert2005}, where the numerical computation of the model's {\sl likelihood function} will be the fundamental concept \re{and main contribution of this work.}

In the research area of eye-movements during reading, a number of competitor models has been proposed. These models implement alternative assumptions on the interaction of word recognition and saccade generation \citep[see][for overviews]{Reichle2003,Rayner2010}. However, there is currently a lack of quantitative model evaluations using objective concepts. First, due to the number of different effects in experimental data, models were often compared qualitatively: Does the model reproduce an experimentally-observed effect or not? Second, in complex cognitive models, parameters were mostly hand-selected or fitted based on minimization of an arbitrary loss-function that quantifies the difference between experimental and simulated data. Third, typical models could not be fitted to data from individual subjects so far. However, explaining interindividual differences is an important aspect of model evaluation, which is precluded when fitting procedures are data hungry and require pooling of data over a large number of participants. Since model identification and model comparison are general problems in psychological and cognitive sciences, \cite{Schuett2017} recently proposed a likelihood-based, statistically well-founded Bayesian framework for parameter estimation in cognitive models. We will demonstrate the feasability of this approach in the case of the SWIFT model for eye-movement control during reading. 

In the following, the data assimilation framework will be applied to the SWIFT model of eye guidance in reading. The remaining part of this section consists of a short introduction to eye movement data and the specifics of likelihood functions for models of fixation sequences.
In Section 2, we describe the details of the SWIFT model. A \re{numerical approximation of the} likelihood function is \re{proposed and tested} in Section 3. In Section 4, we use data from a set of readers to estimate SWIFT parameters and to model their interindividual differences. We close with a discussion \re{of our results} in Section 5.

\subsection{Eye-movement control during reading}
Reading is based on successful word recognition, however, processing of words requires high-acuity vision that is confined to the center of the visual field (the fovea). Therefore, gaze shifts via fast eye movements (saccades) need to be generated to move words into the fovea for word identification. From this general behavioral pattern, reading may be looked upon as an important example of {\sl active vision} \citep{Findlay2003}, which is the notion that eye movements form an essential component for almost all visual perception.

When we read texts, we perform 3 to 4 saccades per second, resulting in fixations on different words with durations between 150 and 300~ms, on average. An example is presented in Figure~\ref{fig_read_example}, where \re{11} fixations are placed on the words of a given sentence. Fixation durations range from \re{110~ms} to 325~ms. In this example, some words are fixated more than once. In the case of an immediate second saccade to the same word as the currently fixated word, the event is called a {\sl refixation} (e.g., fixations 3, a forward refixation, and 5, a backward refixation). Some words are not fixated during first-pass reading, corresponding saccades are termed {\sl skippings} (e.g., word 6, the article ``den'', was skipped \re{in {\sl first-pass} reading).} Furthermore, it happens in roughly 5 to 10\% of the fixations that a corresponding saccade returns to a previously passed region of text, which are called {\sl regressions} (e.g., when word 6, the previously skipped article, receives fixation 9). Taken together, only about 50\% of the saccades are moving the gaze forward from word $n$ to the next word $n+1$, which generates complicated {\sl scanpaths} that are difficult to reproduce and predict by theoretical models of eye guidance during reading.

\begin{figure}[!t]
\unitlength1mm
\begin{picture}(120,30)
\put(0,0){\includegraphics[width=170mm,angle=-0.5]{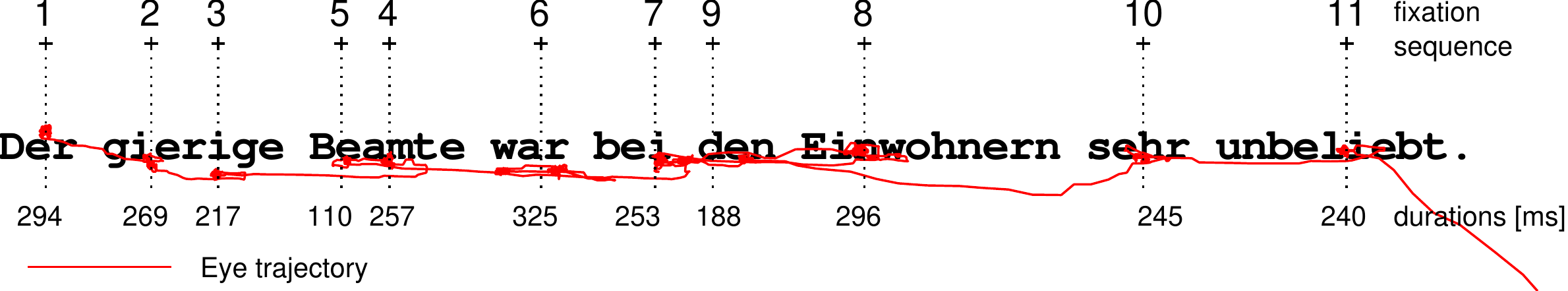}}
\end{picture}
\caption{\label{fig_read_example}
Sequence of fixations during reading. The scanpath indicates a series of fixations and saccades. Fixations are labeled by numbered dotted lines which indicate the horizontal positions. Numbers below the vertical lines are the corresponding fixation durations.
}
\end{figure}

Eye movement research in reading has evolved into one of the fields of cognitive psychology that is strongly driven by computational models. Most of these models are based on simplified assumptions for several cognitive subsystems (e.g., oculomotor circuitry, attention and word recognition), while the core of the models is the orchestration of the subsystems to produce purposeful saccades for reading in a psychologically plausible framework. The way to this success has been paved by the E-Z Reader model \citep{Reichle1998}, a rule-based stochastic automaton model that is based on specific assumptions for the coupling of eye movements and visual attention. This model has been advanced over the years to include more and more specific assumptions \citep[\re{e.g.,}][]{Reichle2009}.

One of the major differences between existing models lies in the generation of different types of saccades (forward saccades, skippings, refixations and regressions). While some models make explicit assumptions on saccade types or are built to have internal states representing saccade types, an alternative model \re{considered here} is motivated by the dynamical field theory of movement preparation \citep{Amari1977,Erlhagen2002}, which communicates the aspiration to form a general framework for human motor control. The SWIFT\footnote{\underline{S}accade Generation \underline{W}ith \underline{I}nhibition by \underline{F}oveal \underline{T}argets} model \citep{Engbert2002,Engbert2005,Schad2012} provides a coherent theoretical framework for reproducing all types of saccades that are observed during reading. Word processing maps to a distributed activation field that serves as a temporally evolving saccade targeting map. This model will be studied in detail with respect to parameter inference.

Given alternative theoretical models, model fitting and model comparisons will become an increasingly important topic in eye-movement research, as in cognitive science in general. So far, the minimization of ad-hoc statistical loss-functions has been used to obtain estimates for model parameters \citep[e.g.,][]{Reichle1998,Engbert2005}. For example, differences in word-frequency dependent \re{distributions} of fixation durations or skipping probabilities have been implemented as a measure of goodness-of-fit. We will replace these procedures by a Bayesian framework that exploits the likelihood function of the model.

Quantitative measures for eye movements during reading are characterized by strong interindividual differences \citep[e.g.,][]{Risse2014}. However, current computational models of eye-movement control could not reproduce and explain these obvious differences in human performance. It is a key message of the current work that the problem of modeling interindividual differences in reading using complex simulation models can be overcome when a likelihood-based framework of model identification, model parameter estimation, and model comparison is applied. \re{We start with a discussion of the general concept of the} likelihood function for dynamical cognitive models in the next section. \re{The approximative computation of the likelihood function for the SWIFT model, which is the central contribution of the current work, is discussed in Section 3.}

\subsection{The likelihood function for dynamical cognitive models}
\label{sec:likedyn}
The key theoretical concept for the current study is the likelihood function \citep[see][for a tutorial]{Myung2003}, which is a quantitative measure of the plausibility of an observation under the assumption of a specific model $M$. We assume that the model depends on a set of parameters $\boldsymbol\theta$ \re{from parameter space $\boldsymbol\Theta$.} In parameter inference, we are interested in the likelihood of the model parameter values $\boldsymbol\theta$ for model $M$ given the experimental data,  
\begin{equation}\label{eq:LikelihoodDef}
    P_M(\boldsymbol\theta|\mbox{data}) = P_M(\mbox{data}|\boldsymbol\theta) \;,
\end{equation}
where $P_M(\mbox{data}|\boldsymbol\theta)$ is the probability of the data given model $M$ with parameters $\boldsymbol\theta$.

The maximum likelihood estimator $\hat{\boldsymbol\theta}_{ML}$ is obtained by maximization of the likelihood function, i.e.,
\begin{equation}\label{eq:MLE}
    \hat{\boldsymbol\theta}_{ML} = \arg \max_{\boldsymbol\theta\in\Theta}P_M(\boldsymbol\theta|\mbox{data}) \;.
\end{equation}

In mathematical models of \re{eye-movement control,} a model must be evaluated against a sequence of fixations. Thus, the data is a time-ordered sequence of fixations $F=\{f_i\}$, where each fixation $f_i$ is characterized by a position $x_i$ on the line of text, a fixation duration $T_i$, and, depending on the model, also a saccade duration $s_i$ \re{between fixation $i-1$ and fixation $i$.}

In a dynamical model, fixation $f_i=(x_i,T_i,s_i)$ is generated from the sequence of previous fixations $f_1\dots f_{i-1}$ under the control of the set of parameters $\boldsymbol\theta$ and, possibly, influenced by internal degrees of freedom $\boldsymbol\xi$, which will be discussed in Section \ref{sec:swiftlik}. Since fixations are time-ordered, we can factorize the likelihood into a product of all fixations $i=1, 2, ...,n$, which are found in the experimental \re{fixation sequence} $F=\{f_i\}_{i=1}^n$, i.e., 
\begin{eqnarray}
    \label{eq:likefact}
    P_M(\boldsymbol\theta|F) &=&
    P_M(\boldsymbol\theta|f_1,\,f_2,\,\dots,\,f_n)\\ 
    &=& P_M(f_1|\boldsymbol\theta) \prod_{i=2}^{n} P_M(f_i|f_1,\,\dots,\,f_{i-1},\boldsymbol\theta) \;, \nonumber
\end{eqnarray}
where $P_M(f_1|\boldsymbol\theta)$ is the probability of the initial fixation starting at time $t=0$. In typical experimental paradigms, however, this probability is one, since the experimental procedure determines the initial fixation position.

For \re{complex cognitive models}, the likelihood function \re{can often} be computed numerically. \re{If numerical computation of the likelihood function is possible,} we must be able to evaluate the likelihood for a large number of combinations of model parameter values $\boldsymbol\theta$ to find the maximum likelihood estimator, Eq.~(\ref{eq:MLE}), \re{based on a given fixation sequence $F$.}

For the implementation of numerical computations, it is advantageous to compute the log-likelihood, given as
\begin{eqnarray}
    \label{eq_loglik}
    l_M(\boldsymbol\theta|F) &=&
    \log(P_M(\boldsymbol\theta|F))\\ 
    &=& \sum_{i=1}^{n} \log(P_M(f_i|f_1,\,\dots,\,f_{i-1},\boldsymbol\theta)) \;, \nonumber
\end{eqnarray}
which prevents the addition of very small \re{numerical} values that typically occur for some of the additive terms $P_M(f_i|f_1,\,\dots,\,f_{i-1},\boldsymbol\theta)$ for the fixations $f_i$.

If we can compute the log-likelihood $l_M(\boldsymbol\theta|F)$ for model $M$ efficiently using numerical simulation, then it will be possible to apply Bayesian parameter inference \citep[for overviews]{Marin2007,Gelman2013}. In Bayesian inference, we seek to compute the posterior distribution $P(\boldsymbol\theta|F)$ over the parameter vector $\boldsymbol\theta$ after the observation of the fixation sequence $F$. In addition to the likelihood that represents constraints from the experimental data, we specify a prior probability $Q(\boldsymbol\theta)$ that indicates our a-priori knowledge on the model parameters. The posterior distribution is given by 
\begin{equation}
   P(\boldsymbol\theta|F) \propto Q(\boldsymbol\theta)P_M(\boldsymbol\theta|F) \;,
\end{equation}
where the constant of proportionality, which is the normalization constant of the posterior, can be omitted, if Markov Chain Monte Carlo (MCMC) methods are used \citep{Gilks1995,Robert2013}.

So far, we discussed the structure of the likelihood function for a single experimentally observed fixation sequence $F$. In a typical experiment, however, we obtain a set of fixation sequences $F_s$ from a participant who read a corpus of $S$ sentences ($s=1, 2, 3, ...,S$), i.e., the data set $\{F_s\}$ is composed of $S$ fixation sequences. Since fixation sequences are statistically independent observations of the reading process, the numerical computation of the likelihood can be carried out independently for each fixation sequence $F_s$. \re{This statistical independence can be exploited to accelerate computations via parallel evaluations of a large number of fixation sequences, which we will discuss in Section \ref{sec:Likelihood}.}

In summary, the likelihood function for dynamical models of sequential data factorizes as explained in Eq.~(\ref{eq:likefact}), which turns out to be basis for incremental numerical computation. If we implement the computation in an efficient way numerically, then Bayesian parameter inference is available using MCMC methods. Before we discuss and apply the MCMC framework, we introduce the SWIFT model in the next section. In Section \ref{sec:swiftlik}, we present the numerical computation of the likelihood function. The MCMC simulation for Bayesian inference will be discussed in Section \ref{sec:Likelihood}.

\section{The SWIFT model of saccade generation during reading}
\label{sec_SWIFT}
Since word recognition is the key process driving eye movements during reading, a natural assumption is that the time-course of ongoing word processing is closely linked to target selection for saccades. In the SWIFT model, each word is represented by a separate activation variable \textcolor{black}{(lexical activation)}  that is tracking the word's current progress in word recognition. The resulting set of lexical activations determines the probability for saccade target selection (so-called spatial or {\sl where} pathway). Whenever a saccade is prepared, the set of lexical activations provides a flexible mechanism for target selection. As time evolves, the relative activations change, so that a continuous-time representation of the next saccade target exists. 

Fixation times are adjusted to the fixated (foveal) word by an inhibitory mechanism (the temporal or {\sl when} pathway). According to an influential proposal \citep{Findlay1999} the spatial and temporal pathways of saccade generation are \re{partially} independent. The SWIFT model is compatible with this view, in the sense that control of fixation duration and saccade target selection are basically independent, however, interactions exist due to the coupling of both pathways via the set of lexical activations.

\subsection{Saccade target selection and temporal evolution of activations}
Each word $m$ in a sentence of $N_w$ words is represented by a time-dependent activation $a_m(t)$. The activation is \re{initially} increasing during lexical access (word recognition), and \re{later}  decreasing during post-lexical processing. The set of activations $\{a_j(t)\}$, $(j=1, 2, 3, ..., N_w)$ must be built up by parallel processing of words, which is the key assumption that distinguishes SWIFT from other models \cite[e.g.,][]{Engbert2011,Reichle2003}. If a saccade target \re{has to be selected} at time $t$, then the probability $\pi_m(t)$ for target selection of word $m$ is given by the relative activation, i.e.,
\begin{equation}
\label{Eq_TargetSelect}
\pi(m,t) = \frac{\left(a_m(t)\right)^\gamma}{\displaystyle\sum_{j=1}^{N_w} \left(a_j(t)\right)^\gamma}  \;,
\end{equation}
which is normalized as $\sum_{m=1}^{N_w}\pi_m(t)=1$ for all $t>0$. The parameter $\gamma$ introduces a weighting of the set of lexical activations, so that switching between different selection schemes is controlled by a variation of $\gamma$:
\begin{equation}
\pi_m(t) \to \left\{ \begin{array}{lcl}
\mbox{winner-takes-all}&: & \gamma\to\infty \\
\mbox{Luce's choice rule}&: & \gamma=1 \quad.\\ 
\mbox{random selection}&: & \gamma\to 0  
\end{array} \right.
\end{equation}

\begin{figure}[!t]
\unitlength1mm
\begin{picture}(150,120)
\put(-3,0){\includegraphics[width=175mm]{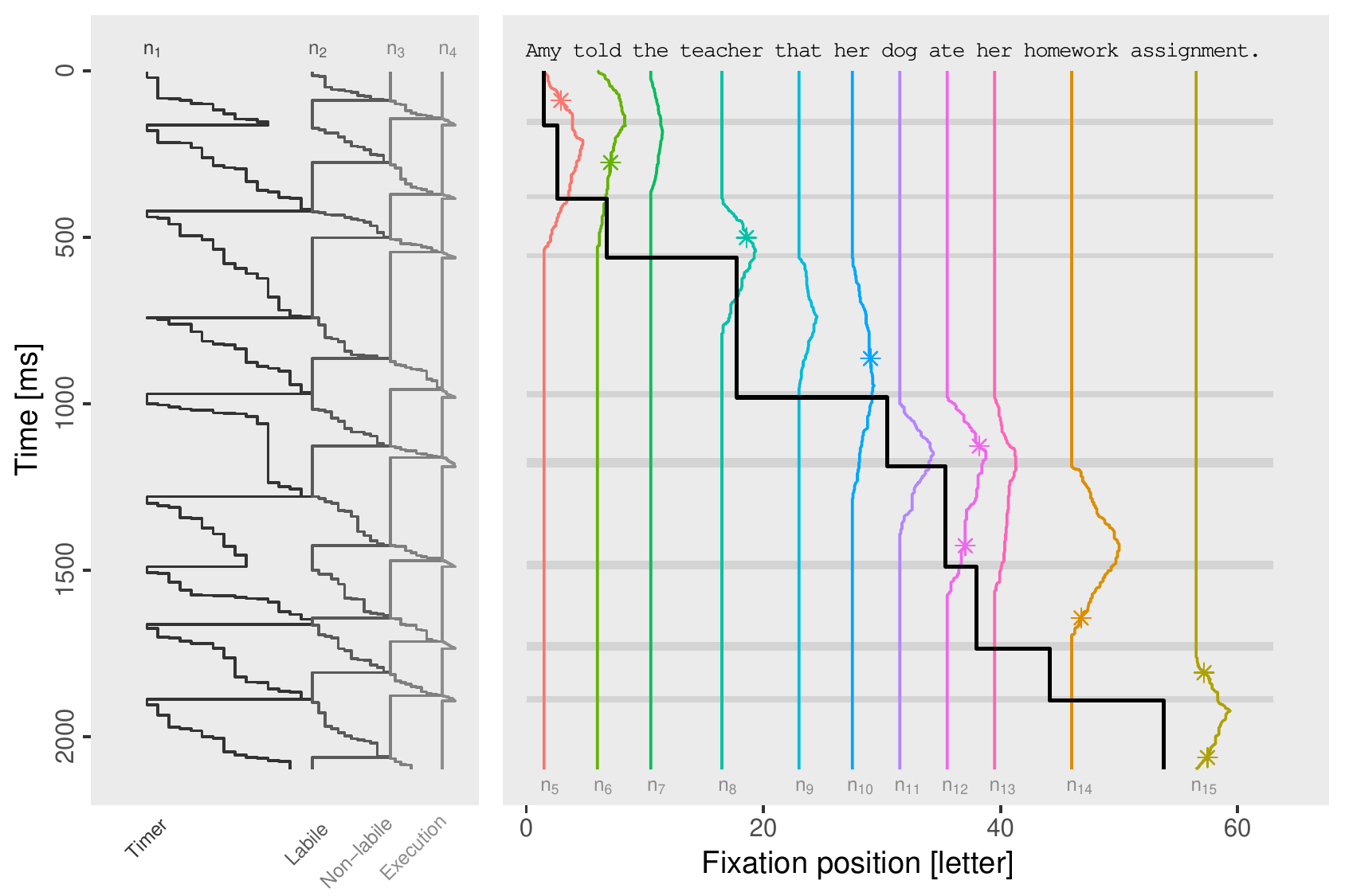}}
\end{picture}
\vspace{-1mm}
\caption{\label{fig_SWIFTeyetrace}
Simulation example for the SWIFT model. \re{The activation field (colored lines) determines the target selection probability $\pi_m(t)$ that evolves dynamically over time (running downwards). The resulting scanpath (fixation sequence) is indicated by the black line. Several random walks (grey, left) generate saccade timer intervals and labile and non-labile saccade latencies. The transition between labile and non-labile stage is the point in time for saccade target selection (asterisk). The saccade timer sends commands to the saccade programming cascade, but also receives inhibition during foveal load (visible shortly after 1000~ms in the example) and is reset for refixations (e.g., second fixation).}}
\end{figure}

An example for a simulated scanpath and the full time-series of lexical activation is illustrated in Figure~\ref{fig_SWIFTeyetrace}. As \re{one can see from} figure, all internal sub-processes of the model are implemented by discrete random walks. In the leftmost column, the saccade timer increases as a one-step process from \re{$n_1=0$} up to a maximum number $N_t$ \re{with transition rate $w_1$.} The stepping rate was chosen as $N_t/t_{\rm sac}$, so that the mean time to reach state $N_t$ is the mean time inter-saccadic time $t_{\rm sac}$ of the model.    

When the saccade timer terminates at state $N_t$, a new saccade timer run is initiated at $n_1=0$ and, at the same time, a labile saccade program start with $n_2=0$ until its threshold $N_l$ is reached. If this labile program terminates, a saccade target is chosen (see asterisks in Fig.~\ref{fig_SWIFTeyetrace}). After the non-labile stage, which is described by state variable $n_3$, the corresponding saccade (state variable $n_4)$ is executed. 

In addition to the saccadic processes, lexical activations are also described by discrete random walks \re{(note, however, the increasing and decreasing parts in the case of lexical activations).} Thus, all sub-processes saccade timing, labile and non-labile saccade programming, saccade execution, and change of lexical activations are represented as one-step \re{stochastic} processes between discrete states.

The state of the model at time $t$ is given by the vector $n=(n_1,\,n_2,\,...,n_{4+N_w})$, where the components $n_j$ represent the states of the subprocesses \re{with transition rates $w_j$. Components} 1 to 4 are saccade-related processes and additional stochastic variables $n_5$ to $n_{4+N_w}$ are keeping track of the (post-)lexical processing of words. We assume a discrete-state, continuous-time stochastic process with Markov property, so that a one-step transition table describes all possible transitions between internal states (Tab.~\ref{Tab_States}). In each of the possible transitions from state $n=(n_1, n_2, ...)$ to $n^\prime=(n^\prime_1, n^\prime_2, ...)$ only one of the \re{components} $n_i$ is changes by one unit, e.g., if the saccade timer generates a transition, then the model's internal change steps from $n=(n_1, n_2, n_3, ...)$ to $n^\prime=(n_1+1, n_2, n_3, ...)$.

\begin{table}[t]\caption{\label{Tab_States}
Stochastic transitions between \re{adjoined} states from $n=(n_1,n_2,...)\mapsto n^\prime=(n^\prime_1,n^\prime_2,...)$}
\vspace{-2ex}
\begin{center}
\begin{tabular}{lcl}
\hline
Process & Transition to ...& Transition rate $W_{n^\prime n}$\\ \hline
Saccade timer      & $n^\prime_1=n_1+1$ & $w_1=N_{t}/t_{sac}\cdot (1+h\,a_k(t)/\alpha)^{-1}$  \\
Labile program     & $n^\prime_2=n_2+1$ & $w_2=N_l/\tau_l$ \\
Non-labile program & $n^\prime_3=n_3+1$ & $w_3=N_n/\tau_{n}$ \\
Saccade execution  & $n^\prime_4=n_4+1$ & $w_4=N_x/\tau_{x}$ \\
Word processing    & $n^\prime_{4+j}=n_{4+j}\pm 1$ & $w_{4+j}=N_a/\alpha\cdot\Lambda_j(t)$ \quad (for word $j$) \\ \hline
\end{tabular}
\end{center}
\end{table}

A numerical algorithm for the simulation of a trajectory of the SWIFT model can be derived easily from our assumptions. The temporal evolution of the probability over the model's internal states is given by the master equation\footnote{The master equation can be interpreted as a conservation equation for probability \citep{Gardiner1985,vanKampen1992}, where the temporal change of probability in state $n$ on the left side of the equation equals the {\sl gain} in probability for state $n$ that is generated by transitions from neighboring states $n^\prime\mapsto n$ and the {\sl loss} in probability generated by transitions from $n$ to neighboring states $n\mapsto n^\prime$.}, 
\begin{equation}
\label{Eq_Master}
\frac{\partial}{{\partial}t}p(n,t|n^{\prime\prime}) = \sum_{n^\prime} \left[ W_{nn^\prime}p(n^\prime,t|n^{\prime\prime}) - W_{n^\prime n}p(n,t|n^{\prime\prime}) \right] \;,
\end{equation}
which is specified by the transition probabilities $W_{n^\prime n}$ for transitions between state vectors $n\mapsto n^\prime$ shown in Table~\ref{Tab_States} with initial condition $p(n^{\prime\prime},0)$, the probability of state $n^{\prime\prime}$ at time $t=0$. When simulating a single trajectory, the system is in a specific state $n$ with certainty and the transition probabilities determine both the waiting time distribution for the next transition and the relative stepping probability to the adjoined states given in (Tab.~\ref{Tab_States}), which will be explained below.

\subsection{Temporal control of saccades and foveal inhibition}
Gaze duration, defined as the sum of the durations of all immediately consecutive fixations on a word, is probably the best measure of required processing time for \re{this} word during natural reading \citep{Rayner1998}. Gaze durations and word recognition times depend linearly on the logarithm of the word's frequency (printed word frequency can be estimated from the word's occurrences in large text corpora). Since word recognition is the basis for text comprehension, an adaptive mechanism for the modulation of fixation duration by word frequency is essential for all models of eye-movement control. 

In general, the required fixation duration for successful word recognition can be attained by two opposing mechanisms: The current fixation can be prolonged by inhibiting the next saccade or, alternatively, the word can be refixated to increase gaze duration. Experimentally, there is only a weak influence of word frequency on the mean first-fixation duration \citep{Kliegl2004}. In contrast, we find a strong effect of word frequency on the probability for refixation. Therefore, there is a preferred strategy for extending the processing time (gaze duration) via generation of a refixation. However, saccade-inhibiting processes can be assumed to contribute a weaker effect (compared to refixation) to the increase in gaze duration by prolonging the ongoing fixation \citep{Engbert2002,Engbert2005}.

Motivated by these observations, the second central assumption in the SWIFT model is {\sl random timing} of fixation duration with additional {\sl foveal inhibition} \citep{Engbert2002} that delays the start of the next saccade program to extend the current fixation.  We assume that foveal inhibition modulates the transition rate $w_1$ for transitions between elementary steps of a random-walk that implements the saccade timer (leftmost column in Fig.~\ref{fig_SWIFTeyetrace}), i.e.,
\begin{equation}
\label{Eq_FovealInhib}
w_1 = \frac{N_t}{t_{\rm sac}}\cdot \left(1 + \frac{h}{\alpha}a_k(t)
\right)^{-1} \;,
\end{equation}
where $N_t$ is the number of states of the timer's random walk and $t_{\rm sac}$ is the mean value of the timer; the activation $a_k(t)$ of the fixated word $k$ (i.e., the word in the fovea) at time $t$ is the key variable that modulates the transition rate of the timer. Using numerical simulations of the model, it can be shown that for $h>0$, foveal inhibition can produce a modulation of the fixation duration that is in good agreement with experimental data \citep{Engbert2002,Engbert2005}. 

\subsection{Character-based visual processing}
Word recognition starts with visual processing of letters, which is done in parallel for all the letters of a given word.  We define the spatial region where word activations can be influenced in the model as the {\sl processing span}.  Within this region, parallel processing is limited by the fact that processing rate depends on the letter's {\sl eccentricity} (i.e., the distance of the letter position from the position of the current fixation). Mathematically, we define an inverted parabolic processing span from the fovea to position $-\delta_L$ on the left and to position $+\delta_R$ on the right of fixation, i.e.,
\begin{equation}
\label{Eq_LettProc}
\lambda(\epsilon) = \lambda_0\cdot\left\{\begin{array}{ccc} 
0\,, & \mbox{for} & \epsilon<-\delta_L \\
1-\epsilon^2/\delta_L^2\,, & \mbox{for} & -\delta_L\le\epsilon<0 \\
1-\epsilon^2/\delta_R^2\,, & \mbox{for} & 0\le\epsilon\le\delta_R \\
0\,, & \mbox{for} & \delta_R<\epsilon \\
\end{array}\right. \;,
\end{equation}
where $\lambda_0$ is a constant given as
\begin{equation}
\lambda_0 = \frac{3}{2}\cdot\frac{1}{\delta_L+\delta_R} \;,
\end{equation}
which is necessary to normalize the total processing rate, i.e., $\int_{-\infty}^{+\infty}\lambda(\epsilon){\rm d}\epsilon=1$.

Experimentally, a strong asymmetry of the {\sl perceptual span} with an extension of 4 to 5 letters to the left of the fixation position and up to 15 letters to the right was found \citep{Rayner1980}.  Therefore, parameters $\delta_L$ and $\delta_R$ should be estimated separately from experimental data. In the following, we estimate $\delta_0\equiv\delta_L=\delta_R$ for simplicity.

\subsection{Word-based processing rate}
Because of the assumption of a processing span, Eq.~(\ref{Eq_LettProc}), processing rates for letters depend on spatial eccentricities. Letter $j$ of word $i$ is processed with rate $\lambda(\epsilon_{ij})$, if it is located at a spatial position with eccentricity $\epsilon_{ij}(t)$ relative to gaze position at time $t$. This letter-based processing rate must be related to the effective word-based processing rate $\Lambda_i(t)$ of word $i$ at time $t$. 

Because of parallel processing of the letters of a given word, each letter contributes to word recognition.  In the case of long words, some letters will have large eccentricities, so that their processing rate will be small (or zero) according to Eq.~(\ref{Eq_LettProc}).  To capture these opposing effects in a parametric model, we make the assumption that the word-based processing rate has the form
\begin{equation}
\label{eq_wordlengthexp}
\Lambda_i(t) = M_i^{-\eta} \sum_{j=1}^{L_i} \lambda(\epsilon_{ij}(t)) \;,
\end{equation}
where \re{$M_i$} is the word length (i.e., number of letters) of word $i$ and $\eta$ is the word length exponent, with $0<\eta<1$.  For $\eta=0$, long words will have a processing advantage. For $\eta=1$, word processing rate is the arithmetic mean of the letter-based processing rates (mean over all letters of a given word); therefore, we will observe a disadvantage for long words in the case $\eta=1$. We expect a numerical value for $\eta$ about 0.5.

With the assumptions on spatial aspects of letter- and word-based processing rates, the temporal aspects of word processing need to be specified.  As discussed for the motivation of the SWIFT model, a time-dependent activation field will provide probabilistic control of saccadic eye movements. Word-based activations $a_i(t)$ for the words of a given sentence are increasing during the initial stage of processing called {\sl lexical processing}. After reaching the maximum of activation $D_i$ for word $i$, the activation starts to decrease ({\sl post-lexical processing}). The maximum of activation is interpreted as processing difficulty, which is a logarithmic function of word frequency $\Omega_i$, i.e.,
\begin{equation}
\label{eq_defdifficulty}
D_i = \alpha\left(1-\beta\frac{\log{\Omega_i}}{\log \Omega^{\rm max}}\right) \;, \end{equation}
where $\Omega^{\rm max}$ is the highest word frequency in a given language and  parameter $\beta$ determines the strength of the word frequency effect.

For word processing, we assume that current activation for each word $i=1,2,3,...,N_w$ is related to the discrete state $n_{4+i}$ of word processing (Tab.~\ref{Tab_States}), given by
\begin{equation}
\label{Eq_ActivationFromStates}
a_i(t) = D_i\frac{n_{4+i}}{N_a} \;,
\end{equation}
where $D_i$\color{black} is the word's processing difficulty, Eq.~(\ref{eq_defdifficulty}). 

{\sl Global decay of activation.}
Maintaining words in working memory during reading cannot be done without loss. Since word activations $\{a_n(t)\}$ represent the state-of-processing, we introduce a global decay of activation. If the processing rate of a word is smaller than the constant $\omega$, then we assume a decay of activation with rate $\omega$.

{\sl Processing during saccades.}
During saccadic eye movements, lexical processing is paused because of {\sl saccadic suppression} \citep{Matin1974}. In the SWIFT model, lexical processing is paused during saccades in the lexical processing stage (increasing activation), while post-lexical processing (decreasing activation) continues during saccades.

\subsection{Oculomotor assumptions}
\label{oculo_assump}
Our assumption of two-stage saccade programming are motivated by the experimental findings of the double-step paradigm \citep{Becker1979}.  A saccade program starts with a labile stage; during this stage, the saccadic gaze center is forced to prepare the next saccade \citep{Findlay1999}, however, a new decision to start a labile saccade program during an ongoing labile stage leads to cancelation and replacement of the earlier saccade program. After the transition to the non-labile stage, the saccade can no longer be canceled or modified. 

Oculomotor errors make an important contribution to eye-movement control during reading. In 1988, based on the analysis of initial fixation positions within words, McConkie and coworkers suggested that a considerable fraction of saccades landed on different words than the intended target words \citep{McConkie1988}. Using an iterative oculomotor modeling approach, \cite{Engbert2008} showed that about 10\% to 20\% of the saccades during natural text reading are mislocated on an unintended word. 

\cite{McConkie1988} showed that saccadic errors can be decomposed into a random (approximately Gaussian) error component and a systematic shift (called saccadic range error). The critical variable that determines the size of both random and systematic error components turned out to be the intended saccade length (distance $d$ from the launch site of the saccade to the center of the target word). Therefore, saccades targeting a word center at $x=0$ will be normally distributed with
\begin{equation}
x\sim {\cal N}(\epsilon_{\rm sre},\sigma^2_{\rm sre}) \;,
\end{equation}
where both parameters depend linearly on the intended saccade length $d$, i.e.,
\begin{eqnarray}
\label{eq:sre}
\epsilon_{\rm sre} &=& r_1-r_2\,d   \\
\label{eq:omn}
\sigma_{\rm sre}   &=& s_1+s_2\,d \;,
\end{eqnarray} 
where $d$ is the physical distance between the launch site of the saccade and the word center of the target word, measured in units of character spaces. The oculomotor parameters $r_1$, $r_2$, $s_1$, and $s_2$ will vary depending on the type of saccade (e.g., refixation or skipping), which is discussed in earlier papers \citep{Engbert2005,Kruegel2010}.
We would like to remark that McConkie et al.'s descriptive model of saccadic errors could be replaced by a process-oriented Bayesian model \citep{Engbert2010,Kruegel2014} in perspective.

\begin{figure*}[!t]
\begin{center}
\unitlength1mm
\begin{picture}(138,51)
\put(5,-5){\includegraphics[width=130mm]{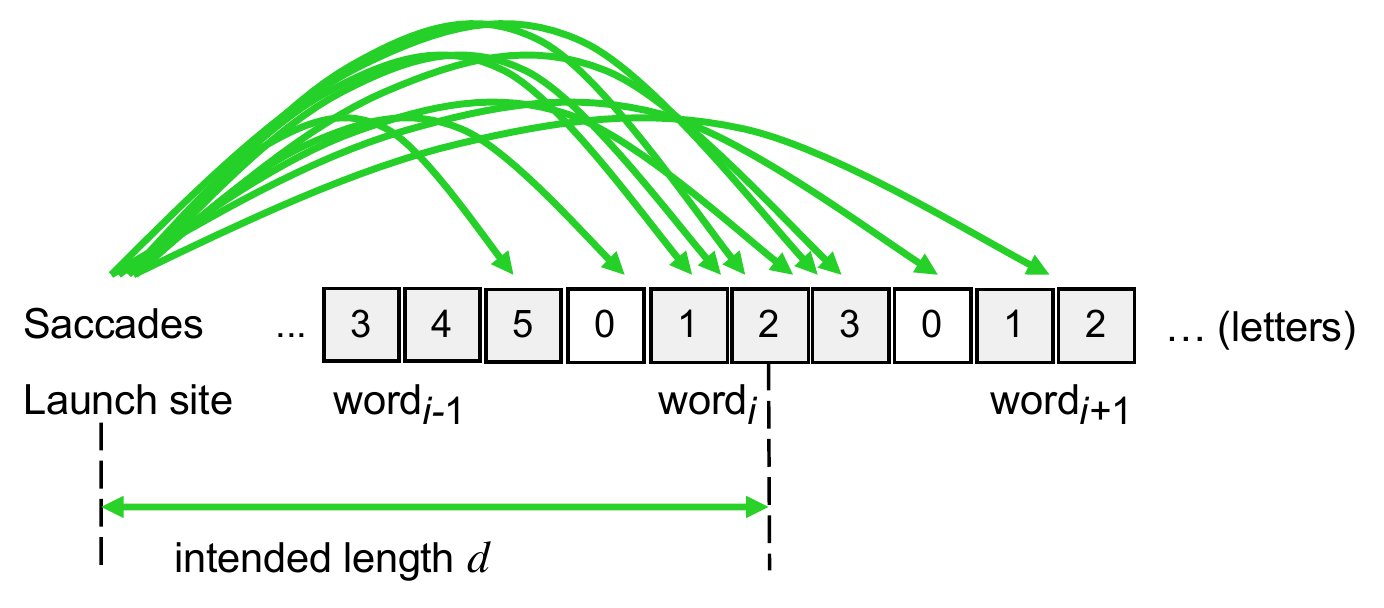}}
\end{picture}
\end{center}
\caption{\label{fig_spatialLik} 
Saccades start at a launch site and aim at the word center of the selected target word $i$. Oculomotor errors are normally distributed, which can lead to misplaced fixations on word $i-1$ (undershoot error) or word $i+1$ (overshoot error). Both the standard deviation $\sigma_{\rm sre}$ and the mean shift $\epsilon_{\rm sre}$ from the intended word's center depend on the intended saccade length $d$. 
}
\end{figure*}

{\sl Modulation of the duration of the labile stage.\label{text_Refixfactor}}
An important problem is the observation of a reduced average fixation duration for refixations. As a solution, we assume that the duration of the labile stage of saccade programming is reduced by factor $R$ ($0<R\le 1$), if the fixation is a refixation. 

Closely related is the phenomenon of mislocated fixations \citep{Engbert2008}. If the realized fixation position (the saccadic landing position) strongly deviates from the word center, so that the landing position will fall onto the neighboring word, then a mislocated fixation will occur. In this case, the duration of the next saccade program will be reduced by factor $M$ ($0<M\le 1$). Such a mechanism is a possible explanation of the inverted optimal viewing-position effect \citep{Nuthmann2005,Vitu2001} of fixation durations that indicates reduced average fixation duration at word edges compared to the word center. \re{In the SWIFT version used here, the probability of misplaced fixation is given as $p_{mis}=0.9\cdot(2\delta/M)^4$, where $\delta$ is the fixation error (distance from word center) and $M$ is the length of the fixated word.}

\subsection{Numerical simulation and model parameters}
For numerical simulations of single trajectories of the SWIFT model, the {\sl minimal process method} by \cite{Gillespie1976}, an exact and efficient numerical algorithm, can be derived from the master equation, Eq.~(\ref{Eq_Master}). If the model is in state $n$ at time $t_0=0$ with certainty, all other states will have zero probability, i.e., $p(n^\prime,t|n)$ for $n^{\prime}\neq n$. Therefore, the master equation, Eq.~(\ref{Eq_Master}), reduces to 
\begin{equation}
\label{eq:Gillespie}
\frac{\partial}{{\partial}t}p(n,t|n) =  -\sum_{n^\prime}W_{n^\prime n}\,p(n,t|n) = -W_n\,\,p(n,t|n) \;,
\end{equation}
where $W_n=\sum_{n^\prime}W_{n^\prime n}$ is the total transition probability from state $n$. From Equation (\ref{eq:Gillespie}), we obtain an exponentially distributed waiting time for the next transition from state $n$ to an adjoined state $n^\prime\neq n$. Following \cite{Gillespie1976}, a two-step algorithm can be derived: In step 1, an exponentially-distributed random number is generated; in step 2, a transition (Tab.~\ref{Tab_States}) is chosen according to relative transition probabilities, $W_{n^\prime n}/W_n$ with $n^\prime\neq n$. This algorithm is numerically efficient, since it restricts computations to the transitions when simulating the system's trajectory. 

For the simulations in this paper we used a restricted version of the SWIFT model to reduce the number of free parameters to 11 \cite[Tab.~\ref{tab_SWIFTpar}; see][]{Engbert2005}. Moreover, we fixed seven of these parameters to estimate four free parameters in the simulation examples. Future simulation studies will be carried out with more free parameters (see Sec.~\ref{sec:Discussion}). The number of possible random-walk states varies between subprocesses; based on earlier simulations \citep{Schad2012}, we used the following numbers: $N_t=15$ (saccade timer), $N_l=12$ (labile saccade stage), $N_n=10$ (non-labile saccade stage), $N_x=20$ (saccade execution), and $N_a=30$ (word activations).

\begin{table}[t]
\caption{\label{tab_SWIFTpar}
Model parameters of the SWIFT model. Numerical values are chosen in agreement with earlier publications (see text).}
\begin{tabular*}{\textwidth}{@{\extracolsep{\fill}}lccc}
\hline
Parameter                      &  Symbol   & Typical Value  & Reference \\ \hline
Lexical difficulty: Intercept  & $\alpha$  & 50  & Eq.~(\ref{eq_defdifficulty}) \\
Lexical difficulty: Slope      & $\beta$   & 0.75   & Eq.~(\ref{eq_defdifficulty}) \\
Processing span        & $\delta_0=\delta_{L,R}$ &  8 & Eq.~(\ref{Eq_LettProc}) \\
Word-length exponent  & $\eta$     & 0.5  & Eq.~(\ref{eq_wordlengthexp}) \\
Saccade timer  & $t_{\rm sac}$  & 250~ms & Tab.~(\ref{Tab_States}) \\
Foveal inhibition    & $h$          & 0.6  & Eq.~(\ref{Eq_FovealInhib}) \\ 
Labile saccade program   & $\tau_l$   & 120~ms  & Tab.~(\ref{Tab_States}) \\
Non-labile program       & $\tau_n$   & 80~ms  & Tab.~(\ref{Tab_States}) \\
Saccade execuation       & $\tau_x$   & 20~ms  & Tab.~(\ref{Tab_States}) \\
Refixation factor        & $R$        & 0.9     & Sec.~\ref{text_Refixfactor} \\
Mislocated fixation      & $M$        & 1.5      & Sec.~\ref{text_Refixfactor} \\
\hline
\end{tabular*}
\end{table}

\section{Likelihood function for the SWIFT model}
\label{sec:swiftlik}
For the parameter estimation procedure discussed in the introduction, we aim at a framework that computes the likelihood of a series of experimentally observed fixations incrementally, Eq.~(\ref{eq:likefact}). For fixation $f_i$, we need to compute the likelihood function $P_M(f_i|f_1,\,\dots,\,f_{i-1},\boldsymbol\theta,\boldsymbol\xi)$ given the previous fixations $f_1, f_2, ..., f_{i-1}$, the model parameters $\boldsymbol\theta$, and the internal states $\boldsymbol\xi$ of model $M$, which we not addressed in Eq.~(\ref{eq:likefact}). In SWIFT each fixation event $f_i=(x_i,T_i,s_i)$ is defined by a fixation position $x_i$ given by the fixated word $v_i$ and the fixated letter $l_i$ within the word}, the fixation duration $T_i$, and the saccade duration $s_i$. The likelihood for fixation $f_i$ is composed of a spatial contribution and a temporal contribution. At time $t$, fixation $i$ starts on letter $l_i$ of word $v_i$, which is predicted by the SWIFT model with a probability determined by word activations and oculomotor assumptions. After fixation $i$ started, the model can make another prediction for the fixation duration $T_i$ of fixation $i$. Next, the likelihood for fixation $i$ can be decomposed into the spatial and temporal contributions, i.e.,  
\begin{equation}
\label{eq:tempspatlike}
P_M(v_i,l_i,T_i|F_{i-1},\boldsymbol\theta,\boldsymbol\xi) = 
P_{\rm temp}(T_i|v_i,l_i,F_{i-1},\boldsymbol\theta,\boldsymbol\xi) 
\cdot P_{\rm spat}(v_i,l_i|F_{i-1},\boldsymbol\theta,\boldsymbol\xi) \;,
\end{equation}
where we introduced $F_{i-1}\equiv\{f_1,\,f_2,\,\dots,\,f_{i-1}\}$ to simplify the notation.

For the {\sl spatial likelihood} $P_{\rm spat}$, the dynamically \re{evolving} word activations in SWIFT determine the time-dependent probability for selecting a particular word as the next target word. \re{Additionally,} the target-selection probability is modified by oculomotor noise. \re{Due to dynamical dependencies,} we compute the likelihood of an experimentally realized fixation position \re{based on the previous fixations.} However, the internal states $\boldsymbol\xi$ are given by the stochastic dynamics and are, therefore, unknown. In principle, we could integrate over many possible realizations of the internal states $\boldsymbol\xi$, which is, however, time-consuming for the numerical computations. Therefore, we compute $P_{\rm spat}$ for one realization of the internal states $\boldsymbol\xi$, which results in fluctuating numerical values for $P_{\rm spat}$. Thus, instead of integrating out the internal degrees of freedom $\boldsymbol\xi$, we used a pseudo-marginal likelihood \citep{Andrieu2009} and eliminated the dependence on $\boldsymbol\xi$ for the spatial likelihood in Eq.~(\ref{eq:tempspatlike}).

For the {\sl temporal likelihood} $P_{\rm temp}$, SWIFT computations start with a realized fixation position on letter $l_i$ of word $v_i$, however, with internal states $\boldsymbol\xi$. Given this fixation position, the distribution of fixation durations can be predicted by the model. The generated estimate of the likelihood of the experimentally realized fixation duration is approximated by averaging over many realizations of the internal states $\boldsymbol\xi$ (e.g., the internal states of the various saccade programming stages). As a result, both $P_{\rm temp}$ and $P_{\rm spat}$ are random variables, which will be discussed in detail in the next two sections.

\subsection{Spatial likelihood}
In SWIFT, saccadic gaze shifts are generated in two steps: First, a target word is determined in a probabilistic selection process based on relative word activations. Second, after a short delay generated by saccade programming, the saccade is executed with oculomotor errors influenced by the saccadic landing position distribution. These oculomotor errors induce stochastic variability in the within-word fixation position and can also induce mislocated fixations \citep{Engbert2008,Nuthmann2005}, where the realized fixation position is placed on a different word than the selected target.

The combination of activation-based saccadic selection and oculomotor errors generates a non-zero probability for all fixation positions (Fig.~\ref{fig_spatialLik}). The target selection probability $\pi(m,t-\tau_n-\tau_x)$ (see. Eq.~\ref{Eq_TargetSelect}) is the probability of selecting word $m$ as a saccade target for a fixation starting at time $t$. It is important to note that target selection occurs at the time of transition from the labile to the non-labile saccade program, so that the probability $\pi(.)$ for selecting the next target word has to be evaluated with an average time delay $\tau_n+\tau_x$. According to our oculomotor assumptions, the saccadic error generates a probability $q(v,l|m,x_{\rm gaze})$ of fixating word $v$ at letter $l$ given that word $m$ is the selected target word and $x_{\rm gaze}$ is the previous gaze position (or saccade launch site). Thus, the spatial likelihood of an observed saccade starting at time $t_i$ towards letter position $l$ of word $v$ is therefore given by 
\begin{equation}
    P_{\rm spat}(v,l|F_{i-1},\boldsymbol\theta) = \sum_{m=1}^{N_w} \pi(m,t_i-\tau_n-\tau_x)\,q(v,l|m,x_{\rm gaze}) \; ,
\end{equation}
where we dropped the conditional arguments to simplify the notation. Moreover, the time-dependency is now written explicitly, since $t_i$ for the computation of the spatial likelihood of fixation $i$ is given by the sum of fixation durations and saccade durations of the previous fixations in the sequence, $t_i=\sum_{l=1}^{i-1} T_l+s_l$.

The oculomotor system generates systematic and random errors that introduce deviations between the target word's center and the realized fixation position. In SWIFT, we adopt McConkie et al.'s (1988) range-error framework by assuming a Gaussian distribution that is shifted with respect to the target word's center. Thus, the probability of landing at letter $l$ of word $v$, given a target word $m$, is given by
\begin{equation}
q(v,l|m,x_{\rm gaze}) = \frac{1}{\sqrt{2\pi}\sigma_{\rm sre}} \exp\left(-\frac{((v_m+\epsilon_{\rm sre})-x_{n,l})^2}{2\sigma_{\rm sre}^2}\right) \cdot \Delta x \;,
\end{equation}
where $v_m$ is the spatial position of the target word's center, $x_{v,l}$ is the spatial position of the fixated letter $l$ of word $v$, and $\Delta x=1$ is the unit width of a letter. The oculomotor parameters $\epsilon_{\rm sre}(d)$ and $\sigma_{\rm sre}(d)$ of the range-error model specify systematic shift (saccadic range error) and standard deviation of the random error (oculomotor noise), respectively, Eqs.~(\ref{eq:sre}, \ref{eq:omn}); the intended saccade length $d = \|v_m-x_{\rm gaze}\|$ is given as the distance between the target word's center $v_m$ and the fixation position before the saccade $x_{\rm gaze}$.

\subsection{Temporal likelihood}
\re{Because of two-stage saccade programming and due to the fact that fixations are bounded by two saccades in time, SWIFT's fixation durations are given} as linear combinations of realizations of random variables. For the saccade timer and saccade programming stages, resulting durations are gamma-distributed random variables, which are generated by continuous-time discrete-state random walks according to the master equation, Eq.~(\ref{Eq_Master}). 

The saccade timer controls the initiation of the saccade programming cascade with consecutive labile and nonlabile stages and a saccade execution stage. The time interval between the end point of the previous and the beginning of the next saccade execution is defined as the experimentally observed fixation duration. However, the saccade timer is continuously inhibited by word activations. As a consequence, the mean waiting times (the inverse of the transition probabilities) of the elementary steps of the saccade timer's random walk will be time-dependent. Additionally, the mean durations of the labile stages of saccade programming depend on the type of fixation (i.e., whether it is a refixation, a mislocated fixation, or neither of these). Finally, if the saccade timer produces a short interval, then saccade cancelation will be likely, which results in a higher mean value of the predicted fixation duration. 

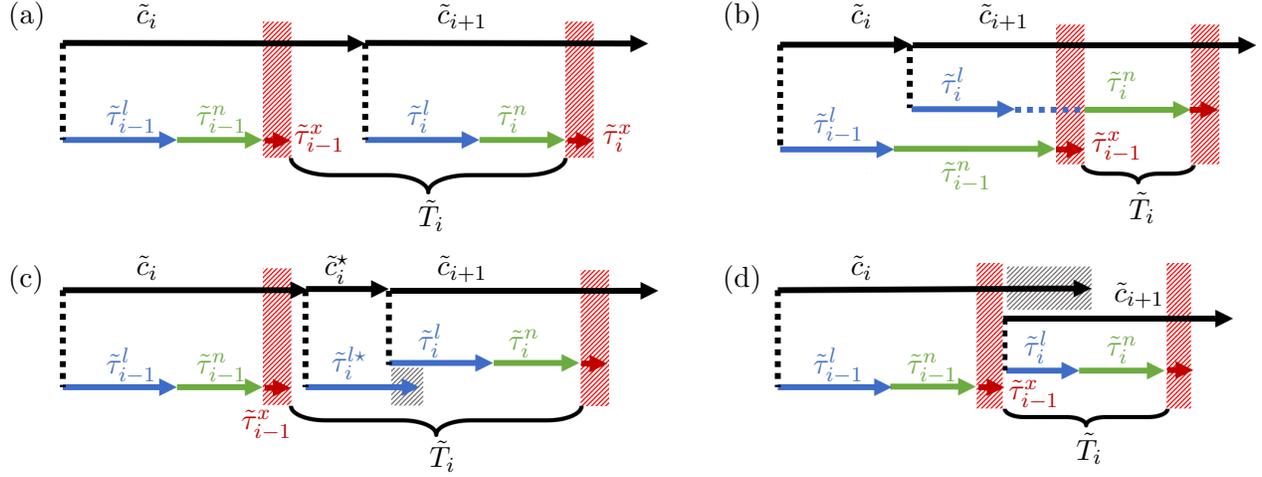
\begin{figure*}[!t]
\begin{center}
\unitlength1mm
\begin{picture}(160,60)
\put(-5,-5){\includegraphics[width=170mm]{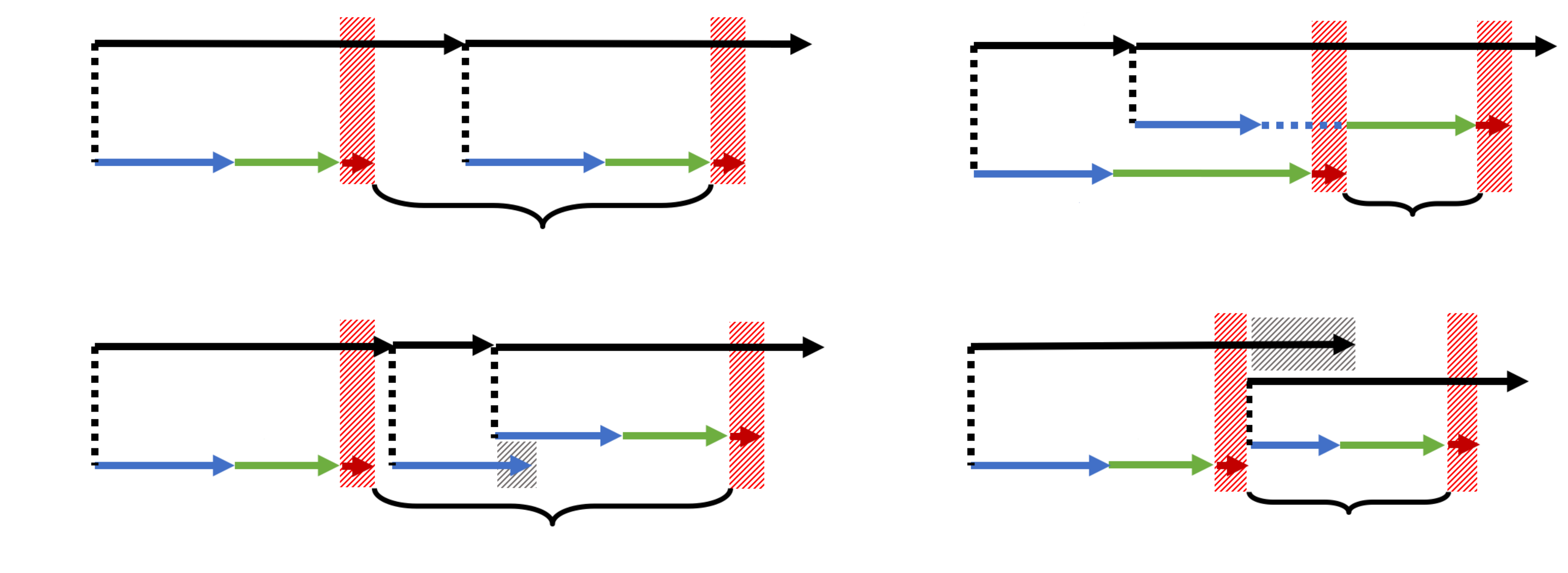}}
\put(-2,55){(a)}
\put(93,55){(b)}
\put(-2,20){(c)}
\put(93,20){(d)}
\put(15,55){$\textcolor{black}{{\tilde{c}}_{i}}$} 
\put(55,55){$\textcolor{black}{{\tilde{c}}_{i+1}}$} 
\put(110,55){$\textcolor{black}{{\tilde{c}}_{i}}$} 
\put(127,55){$\textcolor{black}{{\tilde{c}}_{i+1}}$} 
\put(15,21.5){$\textcolor{black}{{\tilde{c}}_{i}}$} 
\put(40,21.5){$\textcolor{black}{{\tilde{c}}_{i}^{\star}}$} 
\put(55,21.5){$\textcolor{black}{{\tilde{c}}_{i+1}}$} 
\put(110,21.5){$\textcolor{black}{{\tilde{c}}_{i}}$} 
\put(145,18){$\textcolor{black}{{\tilde{c}}_{i+1}}$} 
\put(11,41.5){$\textcolor{PPblue}{\tilde{\tau}_{i-1}^l}$} 
\put(51,41.5){$\textcolor{PPblue}{\tilde{\tau}_{i}^l}$} 
\put(105,40.5){$\textcolor{PPblue}{\tilde{\tau}_{i-1}^l}$} 
\put(122,45.8){$\textcolor{PPblue}{\tilde{\tau}_{i}^l}$} 
\put(11,8.5){$\textcolor{PPblue}{\tilde{\tau}_{i-1}^l}$} 
\put(41,8.7){$\textcolor{PPblue}{\tilde{\tau}_{i}^{l\star}}$} 
\put(52.5,11.7){$\textcolor{PPblue}{\tilde{\tau}_{i}^l}$} 
\put(105,8.5){$\textcolor{PPblue}{\tilde{\tau}_{i-1}^l}$} 
\put(133,10.8){$\textcolor{PPblue}{\tilde{\tau}_{i}^l}$} 
\put(23,41.5){$\textcolor{PPgreen}{\tilde{\tau}_{i-1}^{n}}$} 
\put(63,41.5){$\textcolor{PPgreen}{\tilde{\tau}_{i}^{n}}$} 
\put(122,34){$\textcolor{PPgreen}{\tilde{\tau}_{i-1}^{n}}$} 
\put(144,46){$\textcolor{PPgreen}{\tilde{\tau}_{i}^{n}}$} 
\put(23,8.5){$\textcolor{PPgreen}{\tilde{\tau}_{i-1}^{n}}$} 
\put(64.5,11.7){$\textcolor{PPgreen}{\tilde{\tau}_{i}^{n}}$} 
\put(117.5,8.5){$\textcolor{PPgreen}{\tilde{\tau}_{i-1}^{n}}$} 
\put(144,10.8){$\textcolor{PPgreen}{\tilde{\tau}_{i}^{n}}$} 
\put(36,39){$\textcolor{PPred}{\tilde{\tau}_{i-1}^{x}}$} 
\put(77,39){$\textcolor{PPred}{\tilde{\tau}_{i}^{x}}$} 
\put(142,38){$\textcolor{PPred}{\tilde{\tau}_{i-1}^{x}}$} 
\put(29,1){$\textcolor{PPred}{\tilde{\tau}_{i-1}^{x}}$} 
\put(131,05){$\textcolor{PPred}{\tilde{\tau}_{i-1}^{x}}$} 
\put(52.5,28){$\textcolor{black}{\tilde{T}_i}$} 
\put(147,29){$\textcolor{black}{\tilde{T}_i}$} 
\put(54,-4){$\textcolor{black}{\tilde{T}_i}$} 
\put(140,-3){$\textcolor{black}{\tilde{T}_i}$} 
\end{picture}
\end{center}
\caption{\label{fig_Timers} Schematic illustrations of the generation of fixation durations for different types of fixations in SWIFT.  
(a) {\sl Standard case}: The fixation duration is calculated from the difference of the sum of the saccade timer $c_i$, the labile and nonlabile saccade latencies $\tilde{\tau}^{l}_i$ and $\tilde{\tau}^{n}_i$, respectively, and the sum of saccade latencies $\tilde{\tau}^{l}_{i-1}$, $\tilde{\tau}^{n}_{i-1}$ and $\tilde{\tau}^{l}_{i-1}$. 
(b) {\sl Labile pausing}: If a saccade program reached the non-labile stage it cannot be aborted anymore. A newly started labile programming stage will transition to its non-labile stage only after the current saccade program is terminated at saccade offset.
(c) {\sl Saccade cancelation}: If the saccade timer finishes earlier than the concurrent labile saccade program, the ongoing labile saccade program is canceled---consequently, both the labile program and the saccade timer are restarted. The realized duration of the premature saccade timer $\tilde{c}^*_i$ is added to the new realization $\tilde{c}_i$. 
(d) {\sl Refixation and Mislocated Fixation}: If the current fixation is either a refixation or considered to be a mislocated fixation, the saccade timer realization $\tilde{c}_i$ is reset immediately at fixation onset and a new labile saccade program is initiated. The fixation duration is then given as the sum of the current labile and non-labile durations $\tilde{\tau}^l_{i}$ and $\tilde{\tau}^{n}_{i}$ respectively.} 
\end{figure*}

Since each fixation duration is bounded by two saccades (i.e., the $i$th fixation duration lies between $(i-1)$th saccade offset and $i$th saccade onset), each observed fixation duration $T_i$ is compared to the simulated realization $\tilde{T}_i$ that is given as the sum of the following terms (see Fig.~\ref{fig_Timers}a),
\begin{equation}
\label{Eq_FixDurDef}
    \tilde{T}_i = \tilde{c}_i + \tilde{\tau}^l_i + \tilde{\tau}^{n}_i-\tilde{\tau}^l_{i-1}-\tilde{\tau}^{n}_{i-1}-\tilde{\tau}^{x}_{i-1} \;,
\end{equation}
where $\tilde{c}_i$ is the realized saccade timer duration, $\tilde{\tau}_{i}^l$ and $\tilde{\tau}_{i}^{n}$ are realized durations of the labile and non-labile saccade programming stages respectively, and $\tilde{\tau}_{i}^{x}$ is the realized saccade duration. 

Our strategy for the computation of the temporal likelihood of the $i$th fixation duration $T_i$ is to simulate many realizations of $\tilde{T}_i$ from Eq.~(\ref{Eq_FixDurDef}) \re{to numerically approximate the theoretical distribution of fixation durations with kernel density estimation}\footnote{While it is possible to derive an iterative algorithm for the distribution of linear combinations of gamma-distributed random numbers \citep{Coelho1998,Amari1997}, it turned out that these solutions are numerically unstable.}. \re{In the context of Bayesian analysis, this approach is termed \textit{probability density approximation} (PDA) method \citep{Turner2014,Holmes2015,Palestro2018}, which falls into the broad class of likelihood-free procedures in {\sl approximate Bayesian computation} \citep[ABC; see][for a review]{Sisson2011}.}

Since all of the terms in Eq.~(\ref{Eq_FixDurDef}) are random realizations of stochastic variables, the order of terminations of the subprocesses shown in Fig.~\ref{fig_Timers}(a) can be violated. In the following, we discuss all possible cases:
\begin{enumerate}
\item {\sl Labile pausing} happens if the labile saccade program terminates during an ongoing non-labile saccade program. Since we assume that there cannot be more than one non-labile saccade program active at a time, the current labile program is paused immediately before termination, thus its duration is extended until the current non-labile program and saccade execution finish (Fig.~\ref{fig_Timers}b). Formally, this situation is encountered if $\tilde{c}_i + \tilde{\tau}_i^{l}<\tilde{\tau}_{i-1}^l + \tilde{\tau}_{i-1}^{n}+\tilde{\tau}_{i-1}^{x}$. In this case, the interval $\tilde{\tau}_i^{l}$ is increased and the calculation of $\tilde{T}_i$ is simplified to the duration of the non-labile saccade program, i.e.,
\begin{equation}
   \tilde{T}_i=\tilde{\tau}_i^{n} \;.
\end{equation}
Since the duration of the labile program is extended, however, there will be increased probability for the saccade timer to terminate during the ongoing labile program, while will cause saccade cancelation.

\item {\sl Saccade cancelation} occurs if the main saccade timer realization $\tilde{c}_{i+1}$ terminates during an ongoing labile saccade programming stage $\tilde{\tau}_{i}^l$, i.e.,
$\tilde{c}_i^\star<\tilde{\tau}_i^{l\star}$, which is illustrated in Figure \ref{fig_Timers}c. In this case the labile saccade program is canceled and replaced with the new labile saccade program initiated by restarting of the saccade timer. As a result, the duration of the timer $\tilde{c}_i$ in Eq.~(\ref{Eq_FixDurDef}) is replaced by the sum $\tilde{c}_i+\tilde{c}_i^\star$. Therefore, the corresponding distribution $T_i$ for saccade cancelation is given by
\begin{equation}
\label{Eq_FixDurDef_Cancellation}
    \tilde{T}_i = \tilde{c}_i + \tilde{c}_i^\star + \tilde{\tau}^l_i + \tilde{\tau}^{n}_i-\tilde{\tau}^l_{i-1}-\tilde{\tau}^{n}_{i-1}-\tilde{\tau}^{x}_{i-1} \;,\qquad \mbox{if}\quad c_i^\star<\tilde{\tau}_i^{l\star}\;.
\end{equation}
In principle, saccade cancelation can happen repeatedly within the same fixation, depending on the choice of parameters. 

\item {\sl Refixations and mislocated fixations} represent another special case, where a new saccade program is triggered immediately after the fixation onset (Fig.~\ref{fig_Timers}d). In both cases the saccade timer realization $\tilde{c}_i$ is reset and a new labile saccade program is initiated. The mean duration of the new labile stage is modified by coefficients $f^r=1/R$ and $f^m=1/M$ for refixations and mislocated fixation, resp.~(see~\ref{oculo_assump}). As a result, the observed fixation duration is given as 
\begin{equation}
   \tilde{T}_i = f^{r,m}\tilde{\tau}_i^{l}+\tilde{\tau}_i^{n} \;.
\end{equation}
\end{enumerate}

The SWIFT model includes inhibition of fixation durations by word activation; in its simplest form, the activation of the fixated (foveal) word inhibits the fixation duration by decreasing the transition rates of the saccade timer (Eq.~\ref{Eq_FovealInhib}). Because of the complicated time-course of the activation field (i.e., sudden changes of activation evolution due to saccades), stochastic simulations are necessary to estimate the distribution of $\tilde{T}_i$.

\re{To compute the} likelihood $L_{\rm temp}(T_i)$ of an observed fixation duration $T_i$ \re{we first simulate} the activation evolution for words in the perceptual span from time $t=0$ until the point in time that corresponds to the end of fixation $i$. We start simulating the stochastic contributions by initially going backwards from the time of fixation onset by sampling the saccade latencies $\tilde{\tau}^{x}_{i-1}$, $\tilde{\tau}^{n}_{i-1}$, and $\tilde{\tau}^l_{i-1}$ to determine the onset of the saccade timer $c_i$. The previously sampled activations provide information \re{for} the simulation of the saccade timer with inhibition by foveal word activations, similar to the generative process. If $\tilde{c}_i<\tilde{\tau}^l_{i-1}$, both realizations are discarded and sampled again with the same procedure (we are not interested in saccade cancelation events which do not affect the fixation duration under consideration). The offset of $\tilde{c}_i$ demarks the onset of $\tilde{c}_{i+1}$ and, followin{}g the rules of the previously discussed order violations, we can easily simulate the timer cascade until fixation offset and hence obtain a sample from the distribution of fixation durations as provided by the SWIFT framework with respect to the history of the fixation sequence.

Once $N=300$ fixation durations are sampled, the distribution of $T_i^{\rm exp}$ is approximated via KDE. Increasing the number of samples increases the accuracy of the approximation but is costly in terms of computation time. For the density estimation we use the Epanechnikov kernel \citep{epanechnikov1969} with a bandwidth setting according to Scott's rule \citep{Scott2015}. The Epanechnikov kernel is computationally efficient as it only integrates samples within its limited interval given by the bandwidth. However this can result in situations where no data point is covered by the kernel. To prevent estimates with zero probability, the bandwidth of the kernel was adjusted to the 1.1-fold of the distance between $T_i^{\rm exp}$ and the nearest sample of $\tilde{T}_i$, so that at least one sample will lie within the kernel.

\subsection{Evaluation of the log-likelihood using single-parameter variations}
\label{sec:evalloglik}
A simple test of the likelihood function and its inherent stochastic contributions can be done by repeatedly evaluating the likelihood of a simulated dataset for which the parameters are known and keeping all parameters but one at their respective true values (i.e., the values used in generating the data). Systematically varying the parameter under consideration reveals its impact on the likelihood. Since the likelihood function is composed of two terms from spatial and temporal contributions (Eq.~\ref{eq:tempspatlike}), separating both components can also prove insightful with regard to the strength and direction of the parameter's influence. 

\begin{figure}[!t] 
\unitlength1mm
\begin{picture}(150,160)
\put(-3,80){\includegraphics[width=80mm]{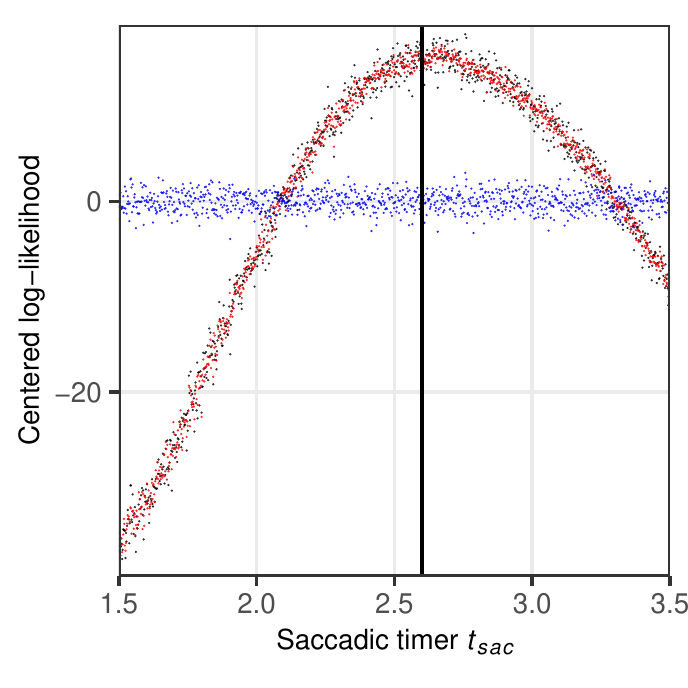}}
\put(82,80){\includegraphics[width=80mm]{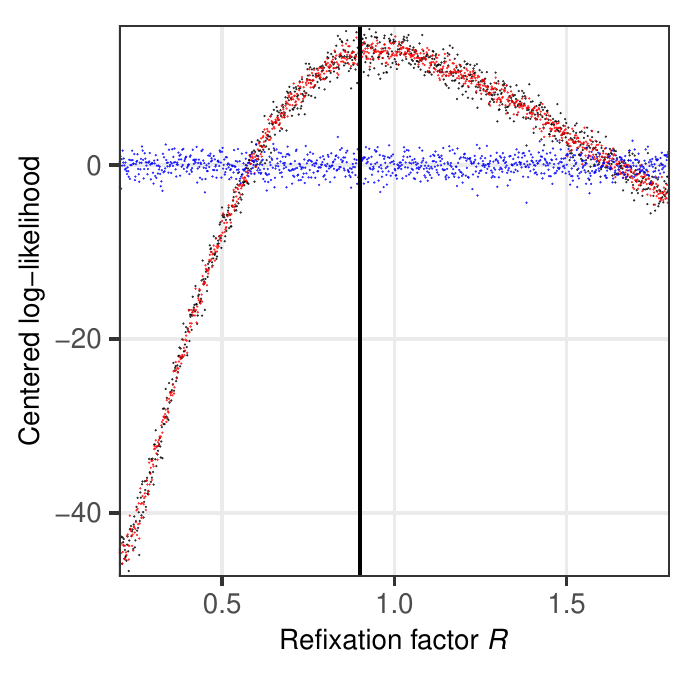}}
\put(-3,0){\includegraphics[width=80mm]{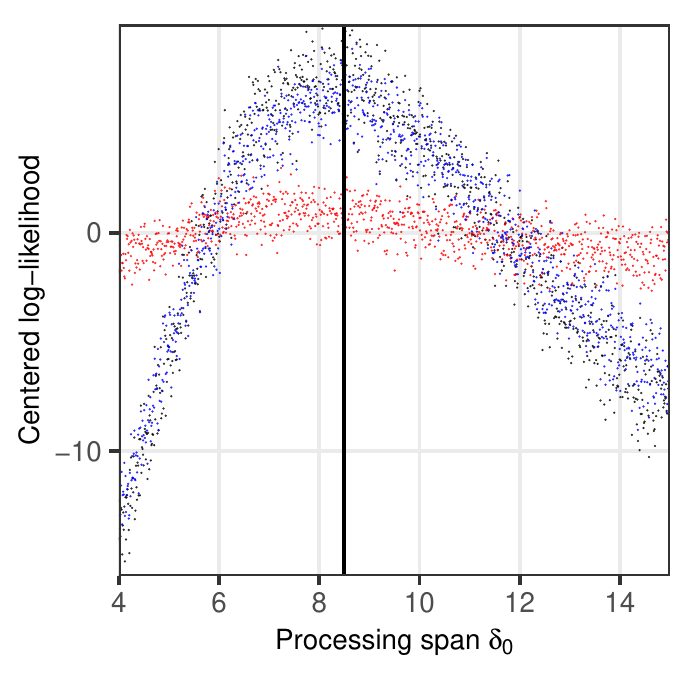}}
\put(82,0){\includegraphics[width=80mm]{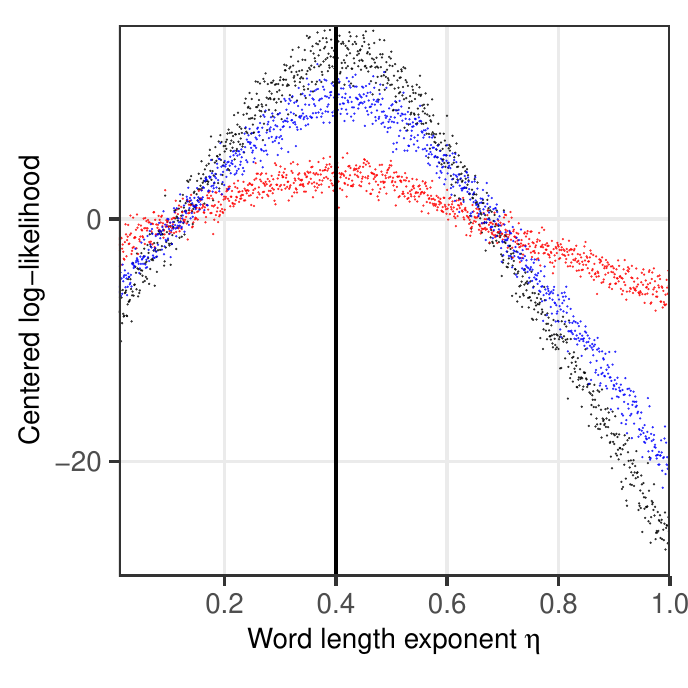}}
\put(60,160){\includegraphics[width=100mm]{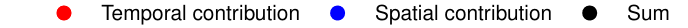}}
\put(1,86){(a)}
\put(86,86){(b)}
\put(1,6){(c)}
\put(86,6){(d)}
\end{picture}
\caption{\label{fig:1dlikefun} Temporal (red) and spatial (blue) contributions to the total (black) log-likelihoods of a simulated dataset (\re{1624} fixations on 114 sentences from the corpus of \cite{RisseSeeligQJEP}). Single parameters were varied within an interval around the respective true parameter value used in creating the data. The log-likelihoods were centered around their respective mean value.} 
\end{figure}
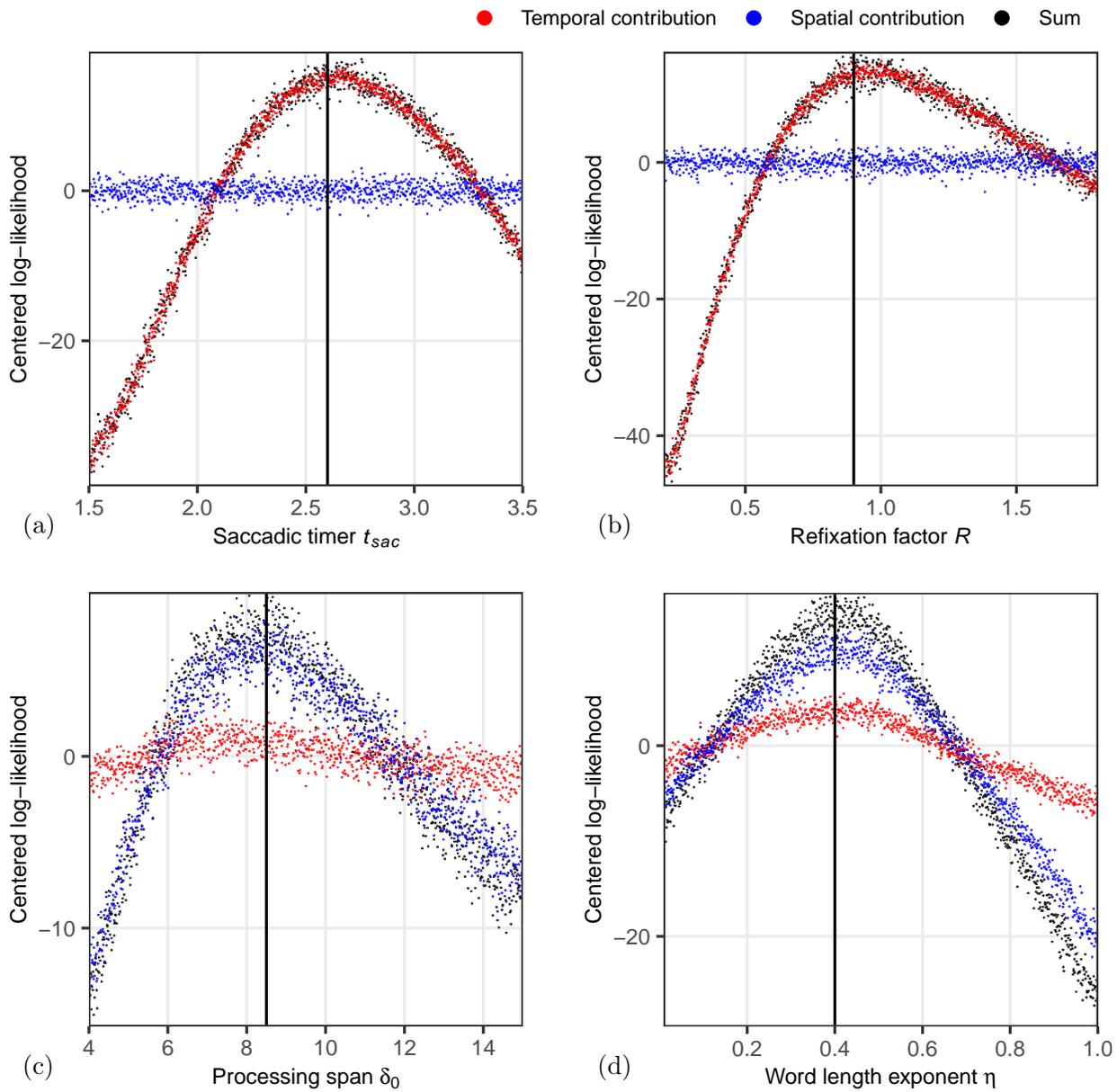

To investigate the properties of the likelihood function for a relevant subset of parameters, we simulated \re{1624} fixations on 114 sentences (Fig.~\ref{fig:1dlikefun}) from the sentence corpus of \cite{RisseSeeligQJEP}. The examined parameters are given in Table~\ref{Tab_ExaminedParams}, with the remaining parameters set according to Table~\ref{tab_SWIFTpar}. The likelihood was then evaluated for 1000 different, evenly spaced values within the given interval (Table~\ref{Tab_ExaminedParams}) separately for each parameter. Since all other parameters were fixed at their true values, any systematic change in the resulting log-likelihood can only be attributed to the parameter under consideration.

\begin{table}[t]\caption{\label{Tab_ExaminedParams}
Parameters of the SWIFT model considered in Bayesian estimation; true values apply to the synthetic data generated for verification of the likelihood function.}
\begin{center}
\begin{tabular}{lccc}

\hline
Parameter            & Symbol     & Range     & True value    \\ \hline
Saccadic timer       & $t_{sac}$  & $150\,...\,350$~ms & $260$~ms      \\
Refixation factor    & $R$        & $0.2\,...\,1.8$    & $0.9$         \\
Processing span      & $\delta_0$ & $4\,...\,15$       & $8.5$         \\
Word length exponent & $\eta$     & $0\,...\,1$        & $0.4$         \\ \hline
\end{tabular}
\end{center}
\end{table}

Figure \ref{fig:1dlikefun}a indicates that the saccade timer $t_{\rm sac}$ influences the temporal likelihood, while there is no influence on the spatial likelihood. A similar behavior is observed for the refixation factor $R$ (Fig.~\ref{fig:1dlikefun}b). In both cases, there is a clear maximum in the likelihood profile at the true parameter values, $t_{\rm sac}=\re{260}$~ms and $R=\re{0.9}$, resp. A different dependence can be seen for the processing span $\delta_0$, which \re{clearly influences} the spatial likelihood (maximum at the true value $\delta_0=\re{8.5}$), but exerts only a minimal influence on the temporal likelihood (Fig.~\ref{fig:1dlikefun}c). For the word-length exponent $\eta$, there is an influence on both spatial and temporal likelihoods (Fig.~\ref{fig:1dlikefun}d), with a maximum for both likelihood profiles at the true parameter value $\eta=\re{0.4}$. 

Thus, our numerical implementation of the likelihood function indicates clear maxima at the true parameter values for simulated data, while stochastic fluctuations due to the approximative account for internal degrees of freedom $\boldsymbol\xi$ are small. In the next section, we will apply an adaptive MCMC framework for Bayesian parameter estimation using simulated and real (experimental) data.

\section{Likelihood-based parameter inference using MCMC}
\label{sec:Likelihood}
With the implementation of the numerical computation of the likelihood function for the SWIFT model from the previous section, we developed the critical step for adopting the Bayesian framework for parameter inference. We will discuss the Markov Chain Monte Carlo approach used for inference, discuss the efficient implementation on a digital computer, present results for parameter recovery from simulated data with known parameters, and, finally, estimate parameters for experimental data.

\subsection{Markov Chain Monte Carlo simulation for the SWIFT model}
\label{sec:MCMC}
As described in Section \ref{sec:likedyn}, the computability of the likelihood $L_M(\boldsymbol\theta|F)$, Eq.~(\ref{eq:likefact}), for a given set of parameters $\boldsymbol\theta$ and a given fixation sequence $F$ is critical for maximum-likelihood and Bayesian inference. For the numerical procedures of Markov Chain Monte Carlo type, we use a variant of the Metropolis Hastings (MH) algorithm \citep{Hastings1970}. In the random-walk MH  algorithm, a random walk in the parameter space is generated, where the probability of the random-walk steps depends on the ratio of the likelihoods associated with the random walk's current and proposed new positions. 

Starting from an arbitrary initial point $X_0$ in parameter space, every move is determined by two steps:
\begin{enumerate}
    \item A proposal $Y_n$ is generated by a random-walk step from  position $X_{n-1}$, 
    \begin{equation}
        Y_n = X_{n-1} + SU_n, 
    \end{equation} where $U_n \sim \mathcal{N}(0,\sigma)$. Both the shape matrix $S$ and the width $\sigma$ of the proposal distribution must be chosen beforehand and kept constant during a run of the algorithm.
    \item The proposal is then accepted with the probability 
    \begin{equation} 
        \alpha_n:=\alpha(X_{n-1},Y_n):=\min\{1,\pi(Y_n)/\pi(X_{n-1})\},
    \end{equation} in which case $X_n = Y_n$, i.e. the walker moves to the proposed position. If the proposal is rejected, then the random walk remains at the current position $X_n = X_{n-1}$.
\end{enumerate}

By recursively following these rules the chain of accepted samples of the algorithm asymptotically converges to the true distribution of $\pi$. However, the speed of convergence greatly depends on an optimal choice of both the shape matrix $S$ and the width parameter $\sigma$ of the proposal distribution. Poor choices lead to abundant rejections (i.e. the chain is stationary most of the time if $S$ is chosen badly or $\sigma$ is too large) or strong autocorrelations of the samples (i.e., movements are very small if $\sigma$ is chosen too small, even if $S$ is optimal). Both parameters are however not known in advance and cannot be obtained due to analytical intractability of SWIFT model's likelihood function. 

Therefore, we used the {\sl Robust Adaptive Metropolis} (RAM) algorithm by \cite{Vihola2012} which progressively captures the parameters' covariance structure shape and at the same time attains a predefined acceptance rate \citep[see][]{Roberts1997}. The speed of the adaptation can also be specified parametrically. Although the RAM algorithm is a good strategy for parameter estimation, it is still computationally expensive, as exploration is naturally slow, if subsequent samples are dependent. Furthermore, it is necessary to use several independent chains with randomly dispersed initial values, each requiring a burn-in phase necessary for the sampler to progress to the vicinity of the stationary distribution. 

An additional modification of the MCMC algorithm is necessary because of the stochastic pseudo-likelihood function of the SWIFT model. If, by chance, an exceptionally high log-likelihood value is obtained for a proposal, the acceptance rate for the subsequent proposal will be very low, \re{which might stall the chain \citep{Holmes2015}}. Therefore we re-evaluate $\pi(X_{n-1})$ for every iteration of the algorithm, which, however, doubles the computation time of the sampling.

To increase computational efficiency, we introduced parallel computation at two levels. First, while the likelihood of a fixation is dependent on all preceding fixations in the respective fixation sequence, likelihoods of whole fixation sequences can be computed independently from each other and added up later. This procedure enables computing the log-likelihood for independent fixation sequences in $F$ in parallel using a multi-core compute cluster. Second, different chains are independent of each other and can therefore be calculated in parallel as well.

\subsection{Parameter recovery using simulated data} 
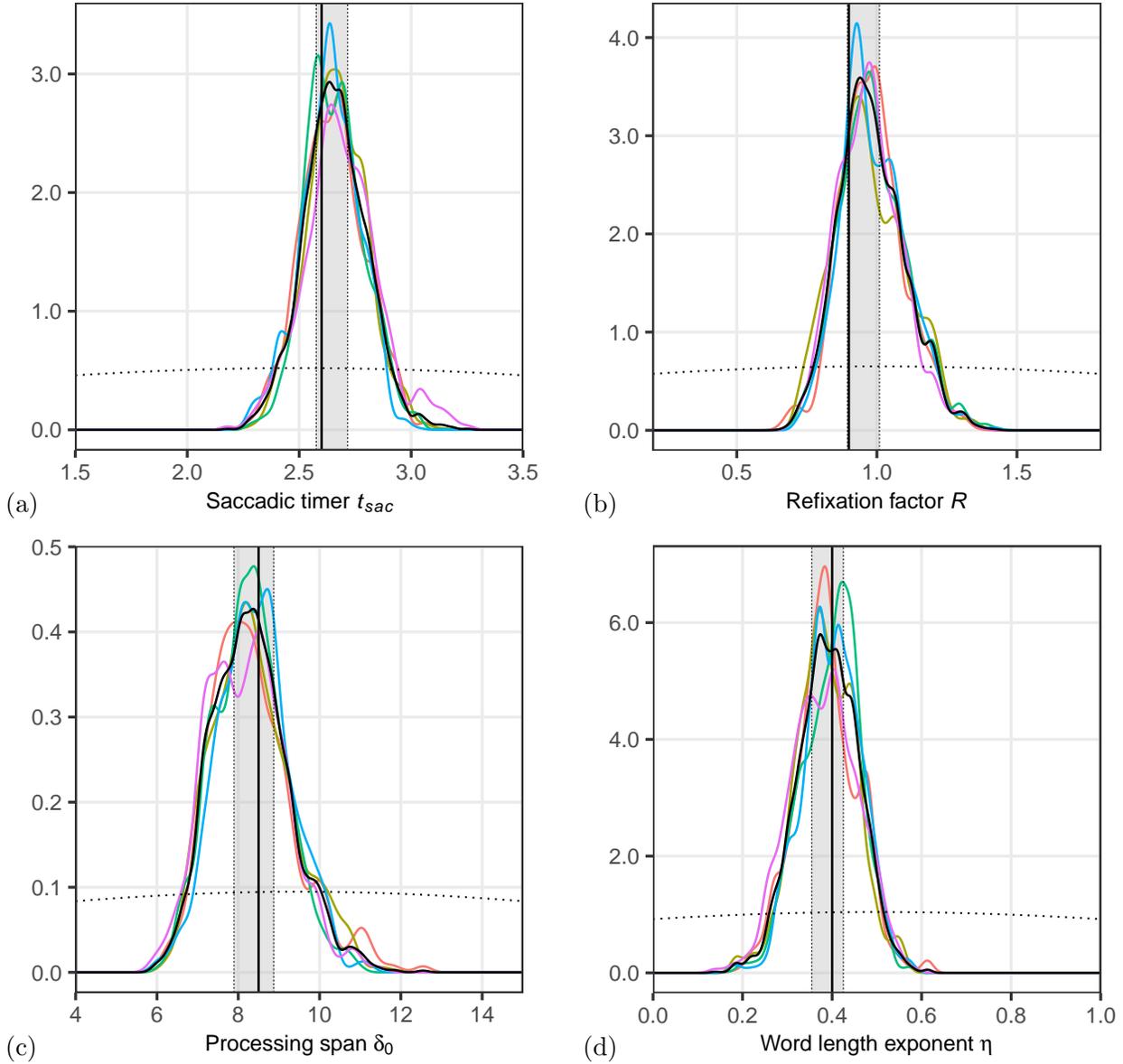
\begin{figure}[!t] 
\unitlength1mm
\begin{picture}(150,150)
\put(0,80){\includegraphics[width=80mm]{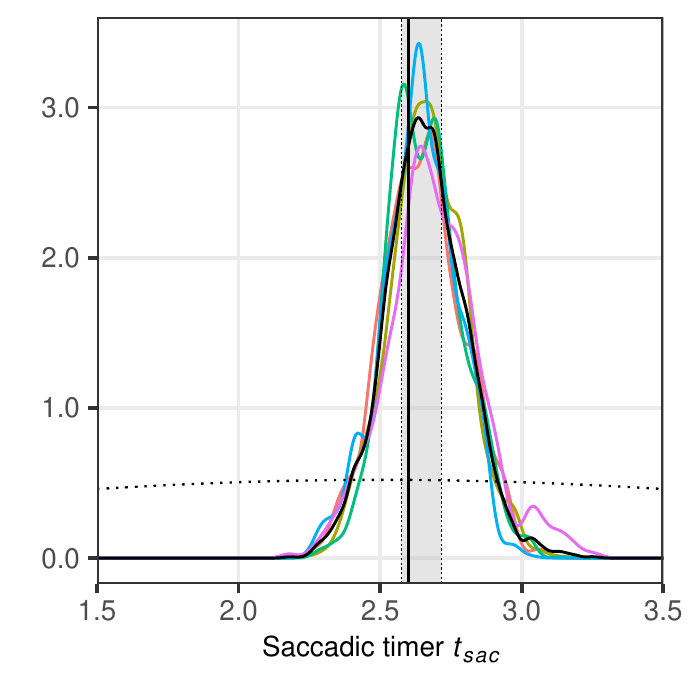}}
\put(85,80){\includegraphics[width=80mm]{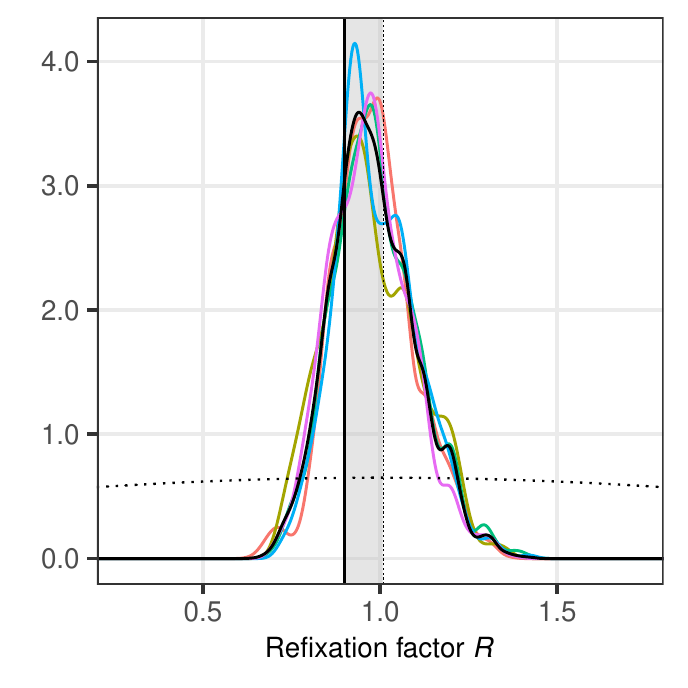}}
\put(0,0){\includegraphics[width=80mm]{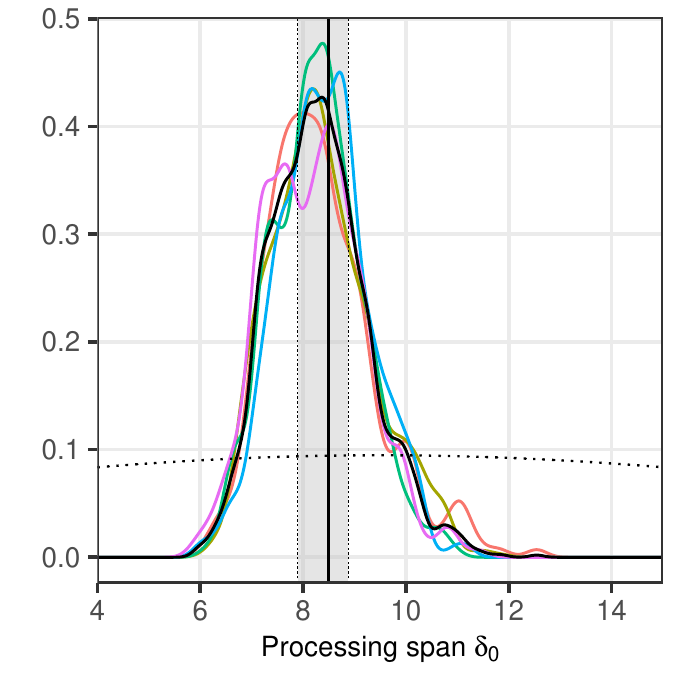}}
\put(85,0){\includegraphics[width=80mm]{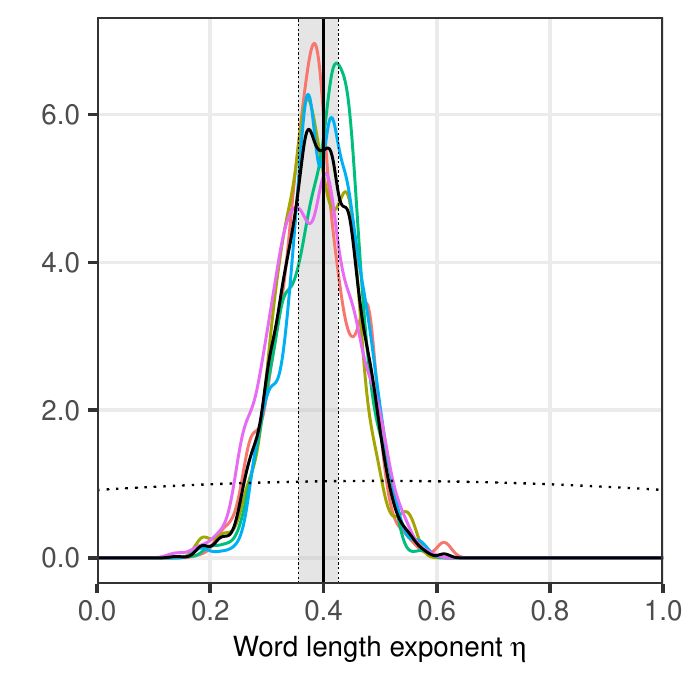}}
\put(1,83){(a)}
\put(86,83){(b)}
\put(1,3){(c)}
\put(86,3){(d)}
\end{picture}
\vspace{-2ex}
\caption{\label{fig:Par_recover} Exemplary Posterior distributions of five individual chains (different colors) for four parameters based on simulated data. The black vertical lines indicate the true parameter values. \re{Grey areas indicate the 40\% HPDI of all chains.} The scale of the parameter range reflects the width of the prior (black, dotted).}  
\end{figure}

Before we \re{demonstrate the application of} the MCMC framework for the SWIFT model to experimental data, we investigate its performance for simulated data with known parameters. While we tested the likelihood function using single-parameter variation around the true value in Section \ref{sec:evalloglik}, we now estimate all four selected parameters \re{(Tab.~\ref{Tab_ExaminedParams})} simultaneously using the MCMC procedure \re{for the same dataset}. We specified truncated normal distributions centered at parameter ranges (see Tab.~\ref{Tab_ExaminedParams}). The standard deviation was set to one half of the estimation range in order to obtain an uninformative prior. We ran 5 independent chains with $N=4,000$ iterations each and the default adaptation parameter value of $\gamma=2/3$. The resulting marginal posterior distributions are given in Figure \re{\ref{fig:Par_recover}, where all true parameter values lie within the 40\% highest posterior density interval (HPDI)}. The results suggest that the likelihood-based MCMC framework is very promising for parameter estimation based on data from single participants.

\subsection{Estimation of parameter\re{s} based on experimental data}
\label{exp_parest}
In the next step, we estimated the same parameters for data from an eye tracking experiment. We used the control condition from a larger experimental study on parafoveal processing using the boundary paradigm \citep[see][for a detailed description of the boundary paradigm]{RisseSeeligQJEP}. We ran 10 chains per participant, each with 4,000 iterations. We used the last 2,000 samples (50\%) after the burn-in to estimate the posterior density. The resulting marginal posterior densities for a single participant are plotted in Figure \ref{fig:parest_w_priors}. While there is an increased variance in the posterior densities for the estimation using experimental data compared to the simulated data \re{(Fig.~\ref{fig:Par_recover})}, we observe clear convergence of the independent chains to a common posterior estimate. Since there is qualitative agreement for the results on simulated and experimental data, the method seems promising to investigate interindividual differences via parameter estimation, which is discussed in the next section.

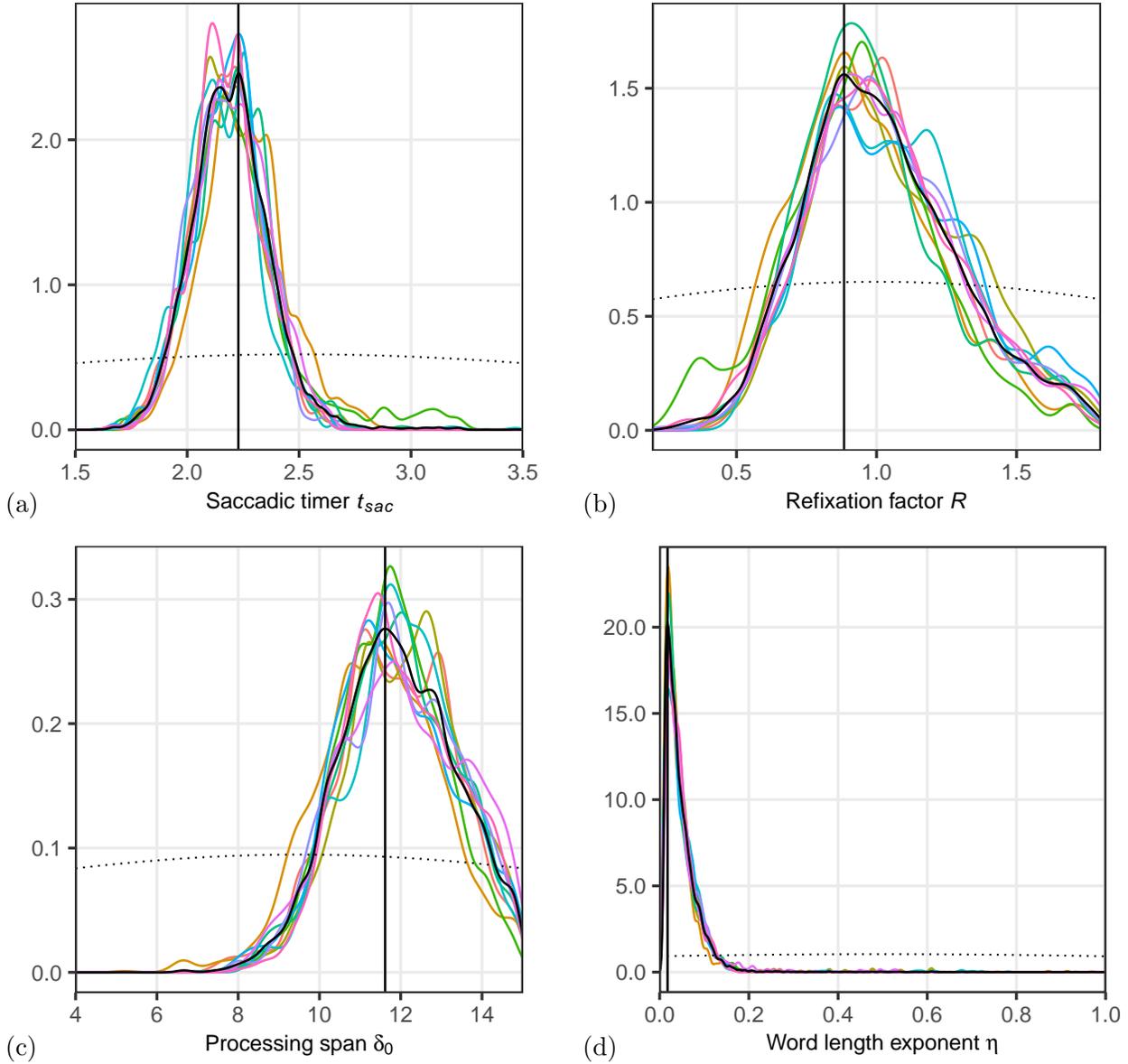
\begin{figure}[!t] 
\unitlength1mm
\begin{picture}(150,150)
\put(0,80){\includegraphics[width=80mm]{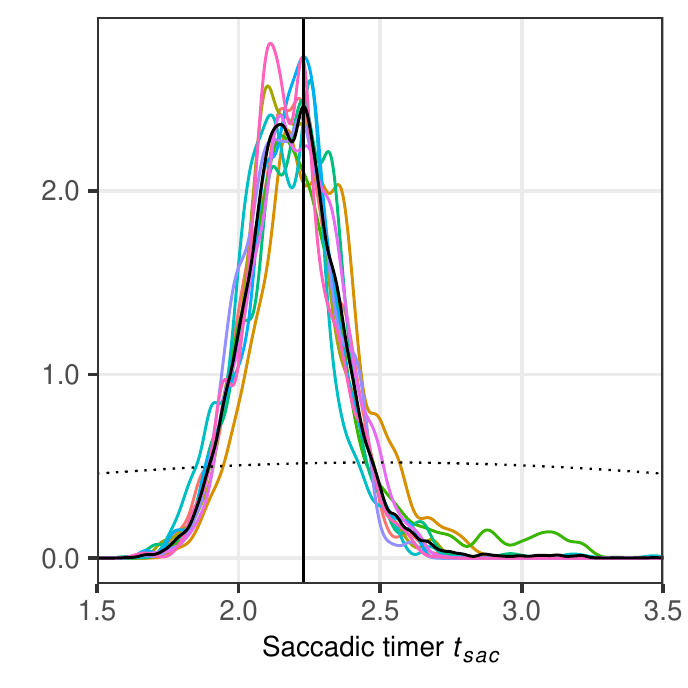}}
\put(85,80){\includegraphics[width=80mm]{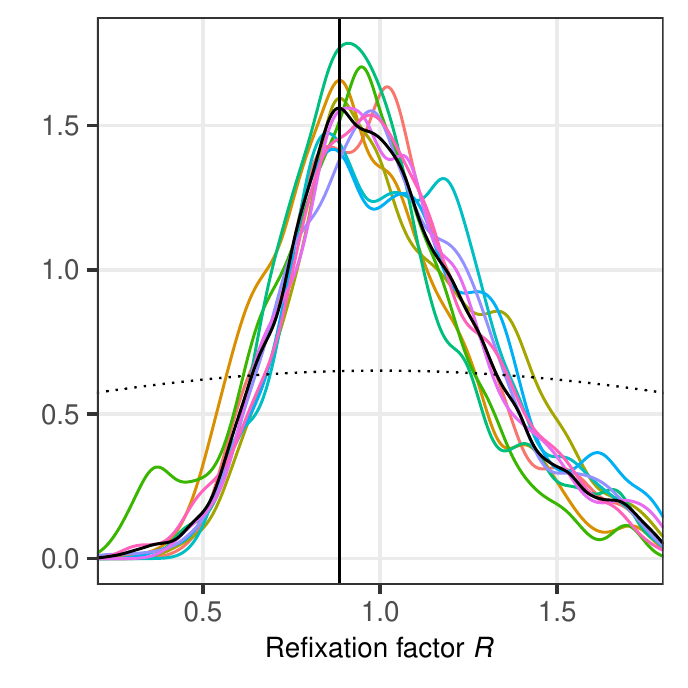}}
\put(0,0){\includegraphics[width=80mm]{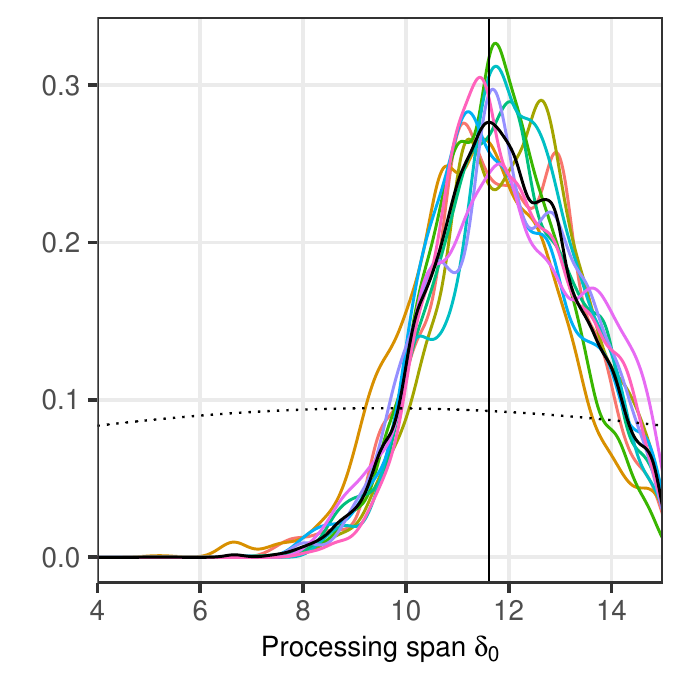}}
\put(85,0){\includegraphics[width=80mm]{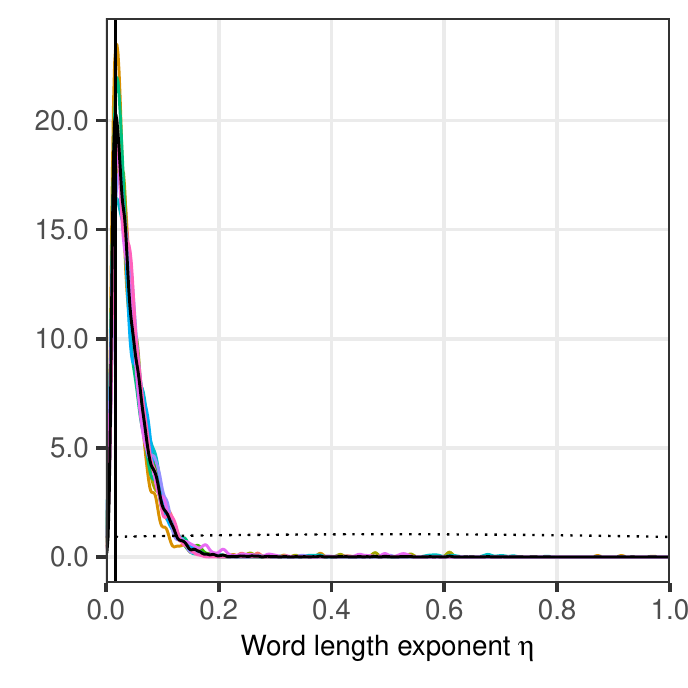}}
\put(1,83){(a)}
\put(86,83){(b)}
\put(1,3){(c)}
\put(86,3){(d)}
\end{picture}
\vspace{-2ex}
\caption{\label{fig:parest_w_priors} Posterior densities for 10 independent chains (coloured) for experimental data from a single participant. The MAP estimator for the pooled chains (black) of each respective parameter is in indicated by the black vertical line. The prior is indicated by the black dotted line.}   
\end{figure}

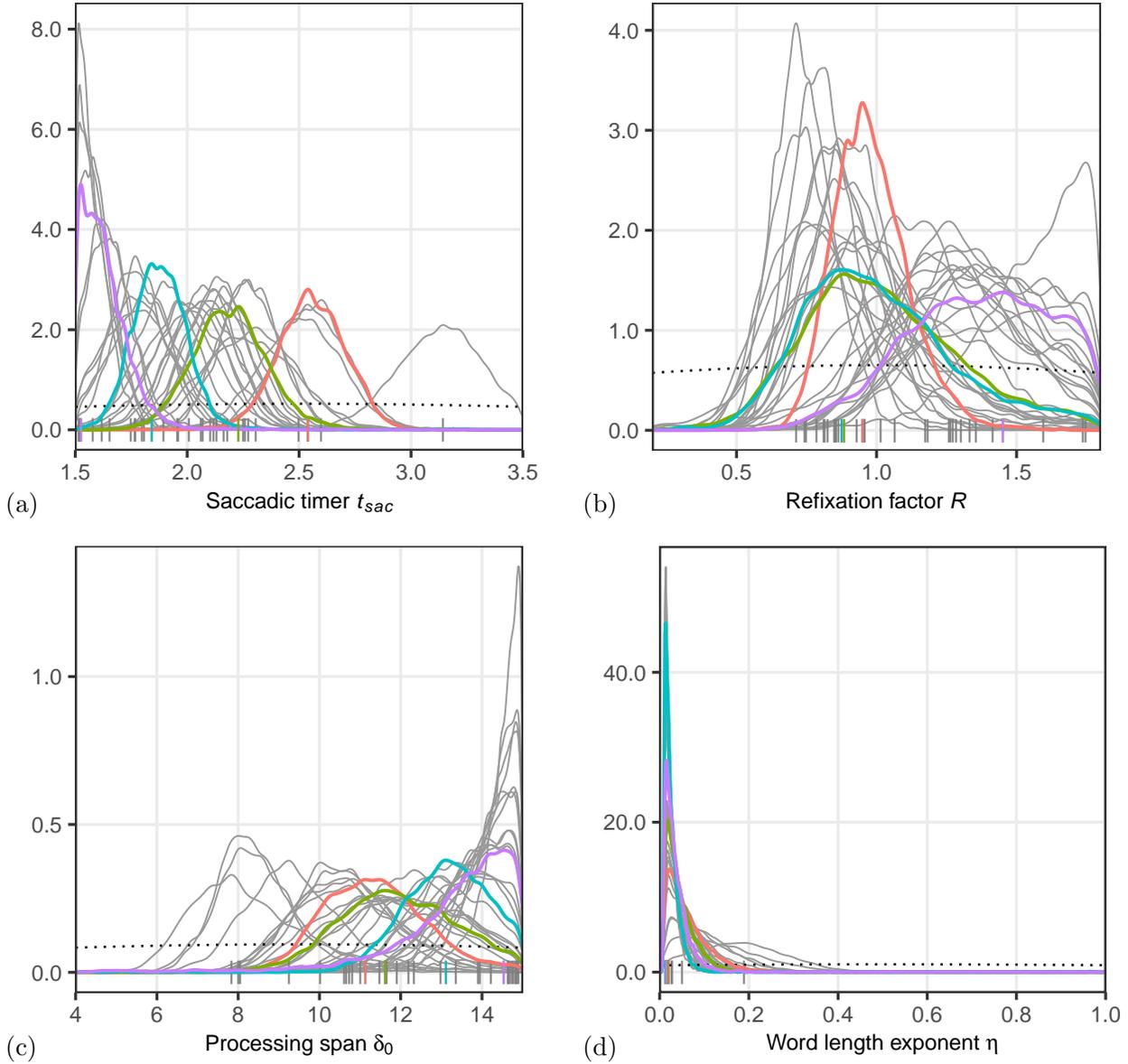
\begin{figure}[!t] 
\unitlength1mm
\begin{picture}(150,150)
\put(0,80){\includegraphics[width=80mm]{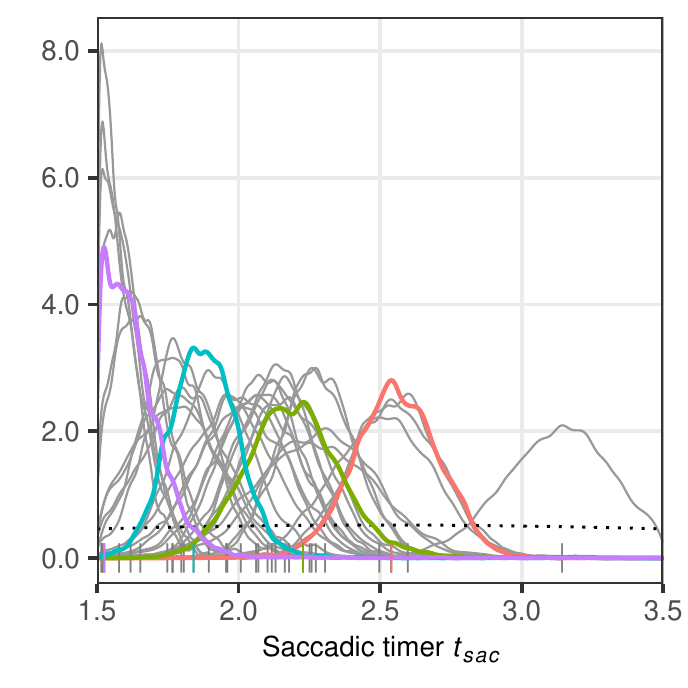}}
\put(85,80){\includegraphics[width=80mm]{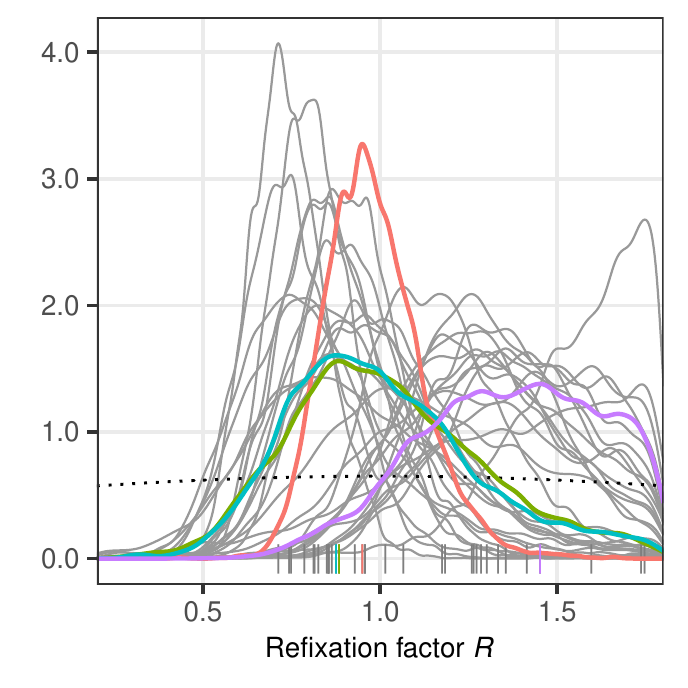}}
\put(0,0){\includegraphics[width=80mm]{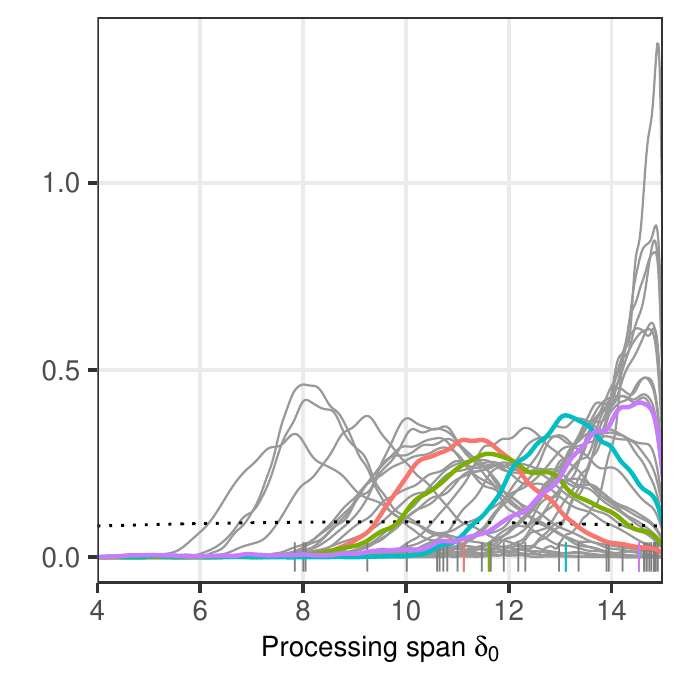}}
\put(85,0){\includegraphics[width=80mm]{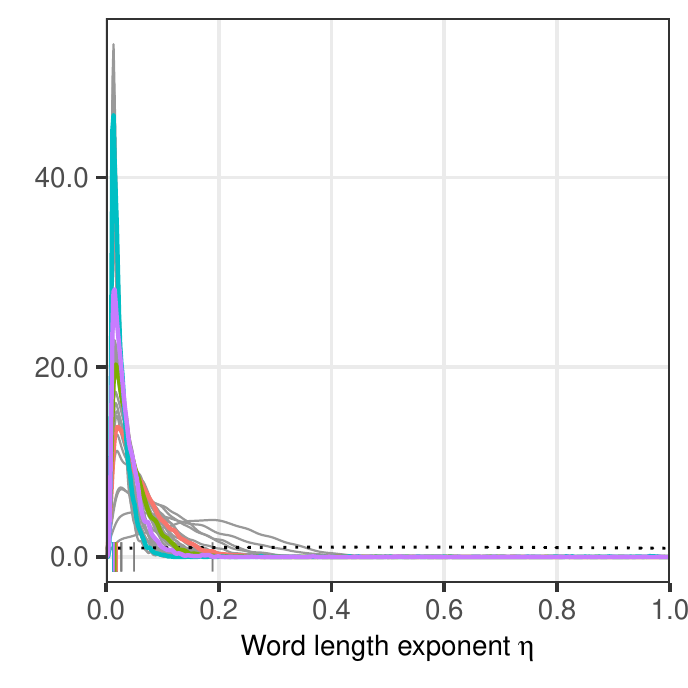}}
\put(1,83){(a)}
\put(86,83){(b)}
\put(1,3){(c)}
\put(86,3){(d)}
\end{picture}
\vspace{-2ex}
\caption{\label{fig:parest_w_priors} Posterior distributions (grey) of 34 participants. Each density is calculated from the pooled data of 10 chains after the burn-in interval. Black ticks at the bottom indicate the MAP estimators for the individual chains. The prior distributions are indicated by the dotted, black line. Curves with the same color correspond to 4 highlighted participants.}   
\end{figure}

\begin{figure}[!t] 
\unitlength1mm
\begin{picture}(150,150)
\put(0,80){\includegraphics[width=80mm]{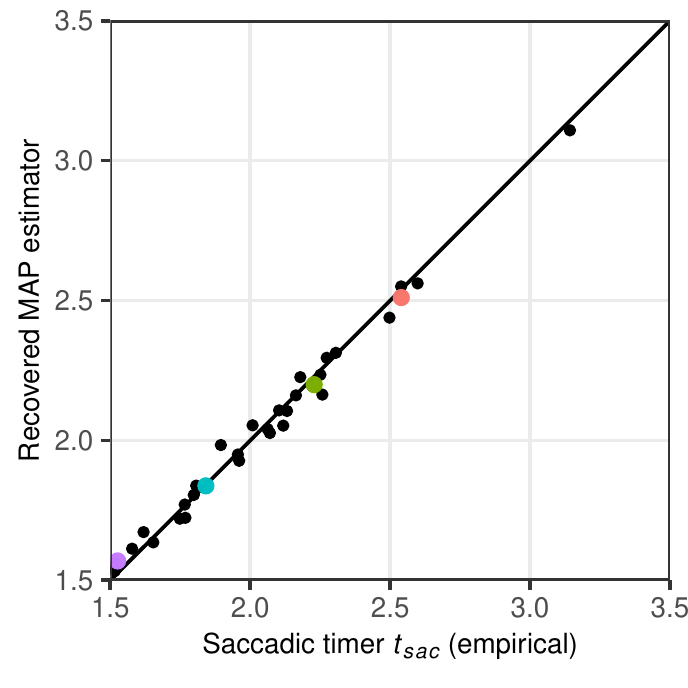}}
\put(85,80){\includegraphics[width=80mm]{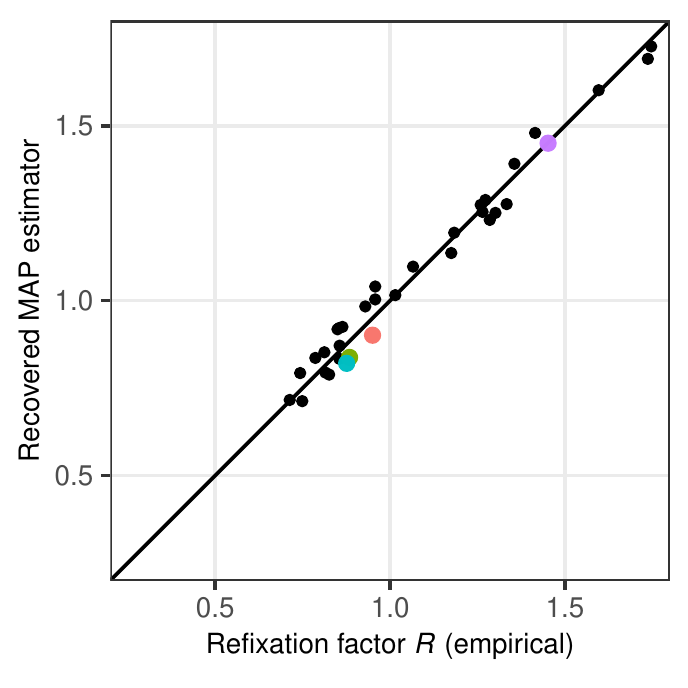}}
\put(0,0){\includegraphics[width=80mm]{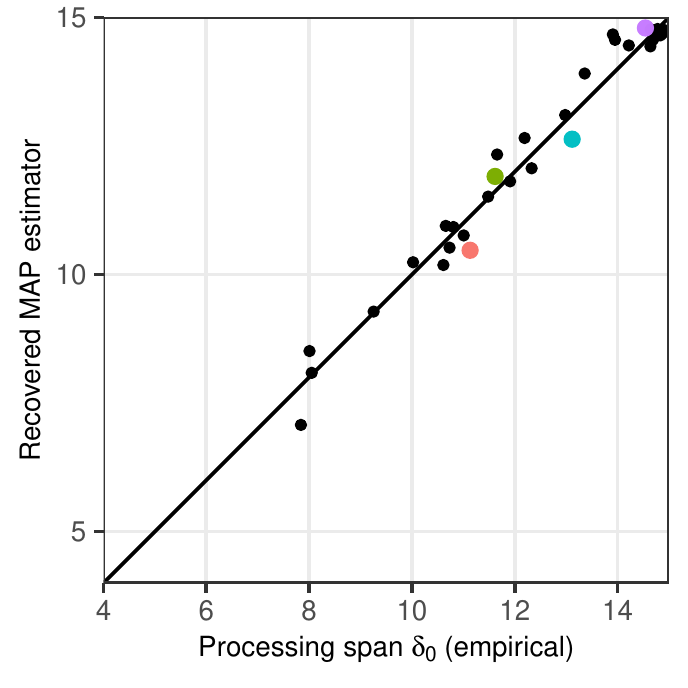}}
\put(85,0){\includegraphics[width=80mm]{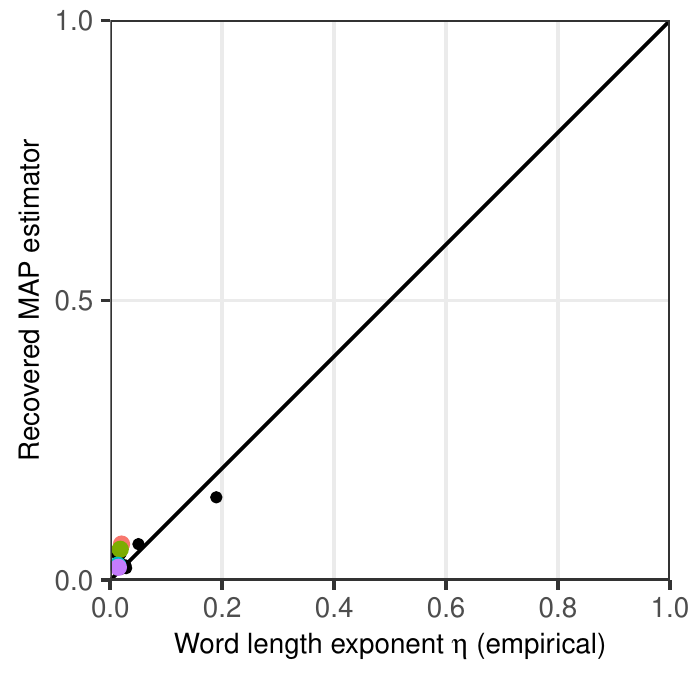}}
\put(1,83){(a)}
\put(86,83){(b)}
\put(1,3){(c)}
\put(86,3){(d)}
\end{picture}
\vspace{-2ex}
\caption{\label{fig:Par_compare_MAP} Relationship between true parameters (horizontal axis) and estimated parameter values of generated data (vertical axis). Parameters used are the MAP estimators for the experimental data. The coloured points correspond to the same participants as in Fig.\ref{fig:parest_w_priors}. }   
\end{figure}

\subsection{Interindividual differences and model parameters}
In this section we study interindividual differences in model parameters across 34 subjects that served as participants in the experiment by \cite{RisseSeeligQJEP}. Figure \ref{fig:parest_w_priors} shows the posterior densities for all subjects, demonstrating considerable interindividual differences over \re{the model parameters $t_{sac}$, $R$, and $\delta_0$, whereas estimates of $\eta$ fall close to zero}.

A critical question is how much of the differences in reading behavior could be explained by the estimated differences in model parameters. Therefore, we used the maximum a posteriori (MAP) estimator (i.e. the mode) of the pooled chains for each subject as input parameters for the generative model and created a simulated data set that corresponds to the experimental data.

\begin{figure}[!t] 
\unitlength1mm
\begin{picture}(150,75)
\put(0,0){\includegraphics[width=80mm]{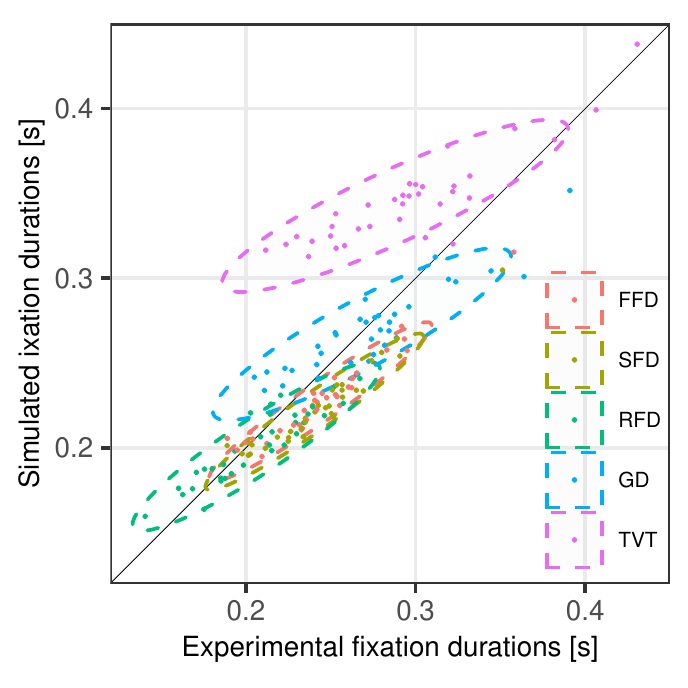}}
\put(85,0){\includegraphics[width=80mm]{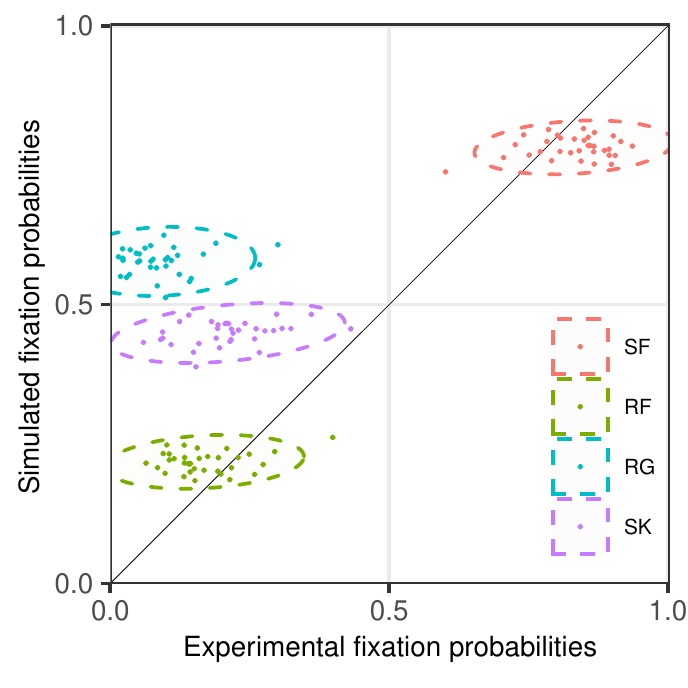}}
\put(1,3){(a)}
\put(86,3){(b)}
\end{picture}
\vspace{-2ex}
\caption{\label{fig:PPwise_comp_fixdurs} (a) Means of different measures of fixation duration for experimental and corresponding simulated data. Each point represents one participant. Simulated data were created using the mean estimated parameters for each respective participant. The coloured ellipses represent the 95\% confidence boundaries. (b) Means of word based fixation probabilities. Again each point represents one participant. }   
\end{figure}

{\sl Fixation durations.} For both the experimental and the artificial data, we calculated participant-wise averages in different measures of fixation durations. Specifically we compared durations of single fixations (\textit{SFD}; when the word was fixated only once in first-pass), first fixations (\textit{FFD}; when the word was fixated once or more in first-pass), refixations (\textit{RFD}; the second fixation on words, which were fixated more than once consecutively in first-pass), gaze durations (\textit{GD}; the total time spent on a word in first-pass) and total viewing time (\textit{TVT}; the total time spent on a word regardless of first, second or more passes). The results (Fig.~\ref{fig:PPwise_comp_fixdurs}a) indicate a remarkably good fit between the experimental data and model predictions for individual participants for \re{RFD and GD. Mean FFD and SFD generated by the model tend to be slightly underestimated for participants with longer initial fixations. Mean TVT, however, is higher in the model predictions than in the experiment. It is important to note that the TVT measure captures more complex gaze behavior, since it also incorporates additional fixation time due to regressions.  }

{\sl Fixation probabilities.} Similar to the analysis of fixation durations, we calculated word-based probabilities for single fixations (SF), refixations (RF), regressions (RG), and word skipping (SK) (Fig.~\ref{fig:PPwise_comp_fixdurs}b). While in the experiment words are more likely to receive single fixations as compared to the simulated data, they consequently have a lower probability of receiving refixations. Additionally, the model predicts higher skipping probabilities and also higher probabilities of serving as regression target. \re{It should be noted that the mismatch between experimental and simulated regression probabilities and experimental and simulated TVT (discussed above) is closely related. In general, part of the regressions might be looked upon as a more complicated psycholinguistic measure related to various aspects of post-lexical processing \citep{Rayner1998} that cannot be captured in the SWIFT model, while another portion of the regressions might be of oculomotor origin and can be found even in scanning tasks \citep{Nuthmann2009}.}

In summary, our results indicate that estimated parameters can explain some of the interindividual differences in fixation durations and fixation probabilities. Thus, the likelihood-based MCMC approach to parameter inference could be applied successfully to estimate model parameters from individual behavioral data.

\section{Discussion}
\label{sec:Discussion}
Current approaches to parameter inference and model comparison \citep[e.g.,][]{Reichle2003} for dynamical cognitive models are insufficient in at least three ways: First, dynamical models need to be tested against time-ordered observations. Second, a likelihood-based procedure is necessary for statistical inference. Third, parameter estimates are needed for individual subjects to explain interindividual differences based on specific model assumptions or components. We set out to solve these three issues in current modeling in computational cognitive science using the SWIFT model of eye-movement control during reading \citep{Engbert2005} as a case study. 
 
The approach discussed here is fundamentally based on the likelihood function of the model. Therefore, we proposed and investigated the numerical likelihood computation of the SWIFT model. This approach is based on the observation that incremental prediction of fixation positions and fixation durations by the generative model can be exploited to determine the likelihood of the next fixation. 

Since the likelihood can be decomposed into a spatial (i.e., fixation position) and a temporal part (i.e., fixation duration), we tried to find separate solutions to both problems. In the spatial part of the likelihood function, internal degrees of freedom (stochastic internal states) could not be integrated out due to numerical efficiency considerations; therefore, we computed a (stochastic) pseudo-likelihood \citep[see][]{Andrieu2009}. In the temporal part, the theoretical likelihood function was unavailable. Therefore, we constructed an approximate likelihood function using a sufficient number of predicted fixation durations from the SWIFT model and KDE for the approximation of the likelihood. In sum, we combined a pseudo-marginal spatial likelihood and an approximated pseudo-likelihood \citep[see][for nomenclature]{Holmes2015} function to obtain the likelihood function of the model \citep{Sisson2011}.

Before we applied our framework to real data, we demonstrated that, in a simplified model version with 4 free parameters, we could reconstruct the true parameter values from simulated data. We used a Bayesian approach using MCMC sampling from the posterior distribution based on an adaptive sampling algorithm \citep{Vihola2012}. The size of the simulated data-set was comparable to a typical experimental data set that is recorded from an individual participant during a one-hour session of eye-tracking experimentation. Next, the same procedure was applied to experimental data. Motivated by the results from simulated data, we estimated model parameters independently for 34 subjects. 

Finally, our results indicate that it is possible to relate interindividual differences in reading behavior (characterized by \re{5} different measures of fixation durations and 4 different measures of fixation probabilities) to differences in the estimated model parameters. Given the \re{typical} state-of-the-art models of eye-movement control in reading, this is a major step for generating hypotheses on the observed interindividual differences in a task as complex as reading. 

Throughout the current work, we focused on the numerical implementation of the likelihood function for the SWIFT model. Since likelihood-based Bayesian inference turned out to be a viable and sound alternative to ad-hoc parameter estimation procedures, we expect that our approach can be further advanced for both theory building and modeling of interindividual differences. \re{For example, for higher dimensional  parameter spaces Differential Evolution MCMC algorithms \citep[see, e.g.,][]{Ter2006,Ter2008,Laloy2012}} might be more adequate. Additionally, we expect that a hierarchical Bayesian design will help to increase the stability of the posterior estimates for individual subjects---even \re{if we apply our methods to data sets smaller than used in the current work.}

\section*{Acknowledgments}
This work was supported by grants from Deutsche Forschungsgemeinschaft (SFB 1294, project B03 to R.E.~and S.Re.; SFB 1287, project B03 to R.E.~and Shravan Vasishth; grant RI 2504/1-1 to S.Ri.). We acknowledge a grant for computing time from Norddeutscher Verbund f\"ur Hoch- und H\"ochstleistungsrechnen (HLRN, grant bbx00001).

\bibliographystyle{apacite}
\bibliography{Likelihood.bib}

\begin{thebibliography}{}

\bibitem [\protect \citeauthoryear {%
S\BHBI i.~Amari%
}{%
S\BHBI i.~Amari%
}{%
{\protect \APACyear {1977}}%
}]{%
Amari1977}
\APACinsertmetastar {%
Amari1977}%
\begin{APACrefauthors}%
Amari, S\BHBI i.%
\end{APACrefauthors}%
\unskip\
\newblock
\APACrefYearMonthDay{1977}{}{}.
\newblock
{\BBOQ}\APACrefatitle {Dynamics of pattern formation in lateral-inhibition type
  neural fields} {Dynamics of pattern formation in lateral-inhibition type
  neural fields}.{\BBCQ}
\newblock
\APACjournalVolNumPages{Biological Cybernetics}{27}{2}{77--87}.
\PrintBackRefs{\CurrentBib}

\bibitem [\protect \citeauthoryear {%
S\BPBI V.~Amari%
\ \BBA {} Misra%
}{%
S\BPBI V.~Amari%
\ \BBA {} Misra%
}{%
{\protect \APACyear {1997}}%
}]{%
Amari1997}
\APACinsertmetastar {%
Amari1997}%
\begin{APACrefauthors}%
Amari, S\BPBI V.%
\BCBT {}\ \BBA {} Misra, R\BPBI B.%
\end{APACrefauthors}%
\unskip\
\newblock
\APACrefYearMonthDay{1997}{}{}.
\newblock
{\BBOQ}\APACrefatitle {Closed-form expressions for distribution of sum of
  exponential random variables} {Closed-form expressions for distribution of
  sum of exponential random variables}.{\BBCQ}
\newblock
\APACjournalVolNumPages{IEEE Transactions on Reliability}{46}{4}{519--522}.
\PrintBackRefs{\CurrentBib}

\bibitem [\protect \citeauthoryear {%
Andrieu%
\ \BBA {} Roberts%
}{%
Andrieu%
\ \BBA {} Roberts%
}{%
{\protect \APACyear {2009}}%
}]{%
Andrieu2009}
\APACinsertmetastar {%
Andrieu2009}%
\begin{APACrefauthors}%
Andrieu, C.%
\BCBT {}\ \BBA {} Roberts, G\BPBI O.%
\end{APACrefauthors}%
\unskip\
\newblock
\APACrefYearMonthDay{2009}{4}{}.
\newblock
{\BBOQ}\APACrefatitle {The pseudo-marginal approach for efficient Monte Carlo
  computations} {The pseudo-marginal approach for efficient monte carlo
  computations}.{\BBCQ}
\newblock
\APACjournalVolNumPages{The Annals of Statistics}{37}{2}{697--725}.
\newblock
\begin{APACrefURL} \url{https://doi.org/10.1214/07-AOS574} \end{APACrefURL}
\newblock
\begin{APACrefDOI} \doi{10.1214/07-AOS574} \end{APACrefDOI}
\PrintBackRefs{\CurrentBib}

\bibitem [\protect \citeauthoryear {%
Becker%
\ \BBA {} J{\"u}rgens%
}{%
Becker%
\ \BBA {} J{\"u}rgens%
}{%
{\protect \APACyear {1979}}%
}]{%
Becker1979}
\APACinsertmetastar {%
Becker1979}%
\begin{APACrefauthors}%
Becker, W.%
\BCBT {}\ \BBA {} J{\"u}rgens, R.%
\end{APACrefauthors}%
\unskip\
\newblock
\APACrefYearMonthDay{1979}{}{}.
\newblock
{\BBOQ}\APACrefatitle {An analysis of the saccadic system by means of double
  step stimuli} {An analysis of the saccadic system by means of double step
  stimuli}.{\BBCQ}
\newblock
\APACjournalVolNumPages{Vision Research}{19}{9}{967--983}.
\PrintBackRefs{\CurrentBib}

\bibitem [\protect \citeauthoryear {%
Beer%
}{%
Beer%
}{%
{\protect \APACyear {2000}}%
}]{%
Beer2000}
\APACinsertmetastar {%
Beer2000}%
\begin{APACrefauthors}%
Beer, R\BPBI D.%
\end{APACrefauthors}%
\unskip\
\newblock
\APACrefYearMonthDay{2000}{}{}.
\newblock
{\BBOQ}\APACrefatitle {Dynamical approaches to cognitive science} {Dynamical
  approaches to cognitive science}.{\BBCQ}
\newblock
\APACjournalVolNumPages{Trends in Cognitive Sciences}{4}{3}{91--99}.
\PrintBackRefs{\CurrentBib}

\bibitem [\protect \citeauthoryear {%
Busemeyer%
\ \BBA {} Townsend%
}{%
Busemeyer%
\ \BBA {} Townsend%
}{%
{\protect \APACyear {1993}}%
}]{%
Busemeyer1993}
\APACinsertmetastar {%
Busemeyer1993}%
\begin{APACrefauthors}%
Busemeyer, J\BPBI R.%
\BCBT {}\ \BBA {} Townsend, J\BPBI T.%
\end{APACrefauthors}%
\unskip\
\newblock
\APACrefYearMonthDay{1993}{}{}.
\newblock
{\BBOQ}\APACrefatitle {Decision field theory: a dynamic-cognitive approach to
  decision making in an uncertain environment.} {Decision field theory: a
  dynamic-cognitive approach to decision making in an uncertain
  environment.}{\BBCQ}
\newblock
\APACjournalVolNumPages{Psychological Review}{100}{3}{432}.
\PrintBackRefs{\CurrentBib}

\bibitem [\protect \citeauthoryear {%
Coelho%
}{%
Coelho%
}{%
{\protect \APACyear {1998}}%
}]{%
Coelho1998}
\APACinsertmetastar {%
Coelho1998}%
\begin{APACrefauthors}%
Coelho, C\BPBI A.%
\end{APACrefauthors}%
\unskip\
\newblock
\APACrefYearMonthDay{1998}{}{}.
\newblock
{\BBOQ}\APACrefatitle {The generalized integer Gamma distribution—a basis for
  distributions in multivariate statistics} {The generalized integer gamma
  distribution—a basis for distributions in multivariate statistics}.{\BBCQ}
\newblock
\APACjournalVolNumPages{Journal of Multivariate Analysis}{64}{1}{86--102}.
\PrintBackRefs{\CurrentBib}

\bibitem [\protect \citeauthoryear {%
Engbert%
\ \BBA {} Kliegl%
}{%
Engbert%
\ \BBA {} Kliegl%
}{%
{\protect \APACyear {2011}}%
}]{%
Engbert2011}
\APACinsertmetastar {%
Engbert2011}%
\begin{APACrefauthors}%
Engbert, R.%
\BCBT {}\ \BBA {} Kliegl, R.%
\end{APACrefauthors}%
\unskip\
\newblock
\APACrefYearMonthDay{2011}{}{}.
\newblock
{\BBOQ}\APACrefatitle {Parallel graded attention models of reading} {Parallel
  graded attention models of reading}.{\BBCQ}
\newblock
\BIn{} S\BPBI P.~Liversedge, I\BPBI D.~Gilchrist\BCBL {}\ \BBA {} S.~Everling\
  (\BEDS), \APACrefbtitle {Oxford {H}andbook of {E}ye {M}ovements} {Oxford
  {H}andbook of {E}ye {M}ovements}\ (\BPGS\ 787--800).
\newblock
\APACaddressPublisher{}{Oxford University Press}.
\PrintBackRefs{\CurrentBib}

\bibitem [\protect \citeauthoryear {%
Engbert%
\ \BBA {} Kr{\"u}gel%
}{%
Engbert%
\ \BBA {} Kr{\"u}gel%
}{%
{\protect \APACyear {2010}}%
}]{%
Engbert2010}
\APACinsertmetastar {%
Engbert2010}%
\begin{APACrefauthors}%
Engbert, R.%
\BCBT {}\ \BBA {} Kr{\"u}gel, A.%
\end{APACrefauthors}%
\unskip\
\newblock
\APACrefYearMonthDay{2010}{}{}.
\newblock
{\BBOQ}\APACrefatitle {Readers use {B}ayesian estimation for eye movement
  control} {Readers use {B}ayesian estimation for eye movement control}.{\BBCQ}
\newblock
\APACjournalVolNumPages{Psychological Science}{21}{3}{366--371}.
\PrintBackRefs{\CurrentBib}

\bibitem [\protect \citeauthoryear {%
Engbert%
, Longtin%
\BCBL {}\ \BBA {} Kliegl%
}{%
Engbert%
\ \protect \BOthers {.}}{%
{\protect \APACyear {2002}}%
}]{%
Engbert2002}
\APACinsertmetastar {%
Engbert2002}%
\begin{APACrefauthors}%
Engbert, R.%
, Longtin, A.%
\BCBL {}\ \BBA {} Kliegl, R.%
\end{APACrefauthors}%
\unskip\
\newblock
\APACrefYearMonthDay{2002}{}{}.
\newblock
{\BBOQ}\APACrefatitle {A dynamical model of saccade generation in reading based
  on spatially distributed lexical processing} {A dynamical model of saccade
  generation in reading based on spatially distributed lexical
  processing}.{\BBCQ}
\newblock
\APACjournalVolNumPages{Vision Research}{42}{5}{621--636}.
\PrintBackRefs{\CurrentBib}

\bibitem [\protect \citeauthoryear {%
Engbert%
\ \BBA {} Nuthmann%
}{%
Engbert%
\ \BBA {} Nuthmann%
}{%
{\protect \APACyear {2008}}%
}]{%
Engbert2008}
\APACinsertmetastar {%
Engbert2008}%
\begin{APACrefauthors}%
Engbert, R.%
\BCBT {}\ \BBA {} Nuthmann, A.%
\end{APACrefauthors}%
\unskip\
\newblock
\APACrefYearMonthDay{2008}{}{}.
\newblock
{\BBOQ}\APACrefatitle {Self-consistent estimation of mislocated fixations
  during reading} {Self-consistent estimation of mislocated fixations during
  reading}.{\BBCQ}
\newblock
\APACjournalVolNumPages{PLoS One}{3}{2}{e1534: 1--6}.
\PrintBackRefs{\CurrentBib}

\bibitem [\protect \citeauthoryear {%
Engbert%
, Nuthmann%
, Richter%
\BCBL {}\ \BBA {} Kliegl%
}{%
Engbert%
\ \protect \BOthers {.}}{%
{\protect \APACyear {2005}}%
}]{%
Engbert2005}
\APACinsertmetastar {%
Engbert2005}%
\begin{APACrefauthors}%
Engbert, R.%
, Nuthmann, A.%
, Richter, E\BPBI M.%
\BCBL {}\ \BBA {} Kliegl, R.%
\end{APACrefauthors}%
\unskip\
\newblock
\APACrefYearMonthDay{2005}{}{}.
\newblock
{\BBOQ}\APACrefatitle {{SWIFT}: {A} dynamical model of saccade generation
  during reading.} {{SWIFT}: {A} dynamical model of saccade generation during
  reading.}{\BBCQ}
\newblock
\APACjournalVolNumPages{Psychological Review}{112}{4}{777--813}.
\PrintBackRefs{\CurrentBib}

\bibitem [\protect \citeauthoryear {%
Epanechnikov%
}{%
Epanechnikov%
}{%
{\protect \APACyear {1969}}%
}]{%
epanechnikov1969}
\APACinsertmetastar {%
epanechnikov1969}%
\begin{APACrefauthors}%
Epanechnikov, V\BPBI A.%
\end{APACrefauthors}%
\unskip\
\newblock
\APACrefYearMonthDay{1969}{}{}.
\newblock
{\BBOQ}\APACrefatitle {Non-parametric estimation of a multivariate probability
  density} {Non-parametric estimation of a multivariate probability
  density}.{\BBCQ}
\newblock
\APACjournalVolNumPages{Theory of Probability \& Its
  Applications}{14}{1}{153--158}.
\PrintBackRefs{\CurrentBib}

\bibitem [\protect \citeauthoryear {%
Erlhagen%
\ \BBA {} Sch{\"o}ner%
}{%
Erlhagen%
\ \BBA {} Sch{\"o}ner%
}{%
{\protect \APACyear {2002}}%
}]{%
Erlhagen2002}
\APACinsertmetastar {%
Erlhagen2002}%
\begin{APACrefauthors}%
Erlhagen, W.%
\BCBT {}\ \BBA {} Sch{\"o}ner, G.%
\end{APACrefauthors}%
\unskip\
\newblock
\APACrefYearMonthDay{2002}{}{}.
\newblock
{\BBOQ}\APACrefatitle {Dynamic field theory of movement preparation.} {Dynamic
  field theory of movement preparation.}{\BBCQ}
\newblock
\APACjournalVolNumPages{Psychological Review}{109}{3}{545}.
\PrintBackRefs{\CurrentBib}

\bibitem [\protect \citeauthoryear {%
Findlay%
\ \BBA {} Gilchrist%
}{%
Findlay%
\ \BBA {} Gilchrist%
}{%
{\protect \APACyear {2003}}%
}]{%
Findlay2003}
\APACinsertmetastar {%
Findlay2003}%
\begin{APACrefauthors}%
Findlay, J\BPBI M.%
\BCBT {}\ \BBA {} Gilchrist, I\BPBI D.%
\end{APACrefauthors}%
\unskip\
\newblock
\APACrefYear{2003}.
\newblock
\APACrefbtitle {Active {V}ision: {T}he {P}sychology of {L}ooking and {S}eeing}
  {Active {V}ision: {T}he {P}sychology of {L}ooking and {S}eeing}\ (\BNUM~37).
\newblock
\APACaddressPublisher{}{Oxford University Press}.
\PrintBackRefs{\CurrentBib}

\bibitem [\protect \citeauthoryear {%
Findlay%
\ \BBA {} Walker%
}{%
Findlay%
\ \BBA {} Walker%
}{%
{\protect \APACyear {1999}}%
}]{%
Findlay1999}
\APACinsertmetastar {%
Findlay1999}%
\begin{APACrefauthors}%
Findlay, J\BPBI M.%
\BCBT {}\ \BBA {} Walker, R.%
\end{APACrefauthors}%
\unskip\
\newblock
\APACrefYearMonthDay{1999}{}{}.
\newblock
{\BBOQ}\APACrefatitle {A model of saccade generation based on parallel
  processing and competitive inhibition} {A model of saccade generation based
  on parallel processing and competitive inhibition}.{\BBCQ}
\newblock
\APACjournalVolNumPages{Behavioral and Brain Sciences}{22}{4}{661--674}.
\PrintBackRefs{\CurrentBib}

\bibitem [\protect \citeauthoryear {%
Gardiner%
}{%
Gardiner%
}{%
{\protect \APACyear {1985}}%
}]{%
Gardiner1985}
\APACinsertmetastar {%
Gardiner1985}%
\begin{APACrefauthors}%
Gardiner, C.%
\end{APACrefauthors}%
\unskip\
\newblock
\APACrefYear{1985}.
\newblock
\APACrefbtitle {Handbook of {S}tochastic {P}rocesses} {Handbook of {S}tochastic
  {P}rocesses}.
\newblock
\APACaddressPublisher{}{Springer-Verlag, New York}.
\PrintBackRefs{\CurrentBib}

\bibitem [\protect \citeauthoryear {%
Gelman%
\ \protect \BOthers {.}}{%
Gelman%
\ \protect \BOthers {.}}{%
{\protect \APACyear {2013}}%
}]{%
Gelman2013}
\APACinsertmetastar {%
Gelman2013}%
\begin{APACrefauthors}%
Gelman, A.%
, Stern, H\BPBI S.%
, Carlin, J\BPBI B.%
, Dunson, D\BPBI B.%
, Vehtari, A.%
\BCBL {}\ \BBA {} Rubin, D\BPBI B.%
\end{APACrefauthors}%
\unskip\
\newblock
\APACrefYear{2013}.
\newblock
\APACrefbtitle {Bayesian {D}ata {A}nalysis} {Bayesian {D}ata {A}nalysis}.
\newblock
\APACaddressPublisher{}{Chapman and Hall/CRC}.
\PrintBackRefs{\CurrentBib}

\bibitem [\protect \citeauthoryear {%
Geyken%
}{%
Geyken%
}{%
{\protect \APACyear {2007}}%
}]{%
Geyken2007}
\APACinsertmetastar {%
Geyken2007}%
\begin{APACrefauthors}%
Geyken, A.%
\end{APACrefauthors}%
\unskip\
\newblock
\APACrefYearMonthDay{2007}{}{}.
\newblock
{\BBOQ}\APACrefatitle {The {DWDS} corpus: {A} reference corpus for the German
  language of the 20th century} {The {DWDS} corpus: {A} reference corpus for
  the german language of the 20th century}.{\BBCQ}
\newblock
\APACjournalVolNumPages{Collocations and idioms: Linguistic, lexicographic, and
  computational aspects}{}{}{23--40}.
\PrintBackRefs{\CurrentBib}

\bibitem [\protect \citeauthoryear {%
Gilks%
, Richardson%
\BCBL {}\ \BBA {} Spiegelhalter%
}{%
Gilks%
\ \protect \BOthers {.}}{%
{\protect \APACyear {1995}}%
}]{%
Gilks1995}
\APACinsertmetastar {%
Gilks1995}%
\begin{APACrefauthors}%
Gilks, W\BPBI R.%
, Richardson, S.%
\BCBL {}\ \BBA {} Spiegelhalter, D.%
\end{APACrefauthors}%
\unskip\
\newblock
\APACrefYear{1995}.
\newblock
\APACrefbtitle {Markov Chain {M}onte {C}arlo in Practice} {Markov chain {M}onte
  {C}arlo in practice}.
\newblock
\APACaddressPublisher{}{Chapman and Hall/CRC}.
\PrintBackRefs{\CurrentBib}

\bibitem [\protect \citeauthoryear {%
Gillespie%
}{%
Gillespie%
}{%
{\protect \APACyear {1976}}%
}]{%
Gillespie1976}
\APACinsertmetastar {%
Gillespie1976}%
\begin{APACrefauthors}%
Gillespie, D\BPBI T.%
\end{APACrefauthors}%
\unskip\
\newblock
\APACrefYearMonthDay{1976}{}{}.
\newblock
{\BBOQ}\APACrefatitle {A general method for numerically simulating the
  stochastic time evolution of coupled chemical reactions} {A general method
  for numerically simulating the stochastic time evolution of coupled chemical
  reactions}.{\BBCQ}
\newblock
\APACjournalVolNumPages{Journal of Computational Physics}{22}{4}{403--434}.
\PrintBackRefs{\CurrentBib}

\bibitem [\protect \citeauthoryear {%
Haken%
, Kelso%
\BCBL {}\ \BBA {} Bunz%
}{%
Haken%
\ \protect \BOthers {.}}{%
{\protect \APACyear {1985}}%
}]{%
Haken1985}
\APACinsertmetastar {%
Haken1985}%
\begin{APACrefauthors}%
Haken, H.%
, Kelso, J\BPBI S.%
\BCBL {}\ \BBA {} Bunz, H.%
\end{APACrefauthors}%
\unskip\
\newblock
\APACrefYearMonthDay{1985}{}{}.
\newblock
{\BBOQ}\APACrefatitle {A theoretical model of phase transitions in human hand
  movements} {A theoretical model of phase transitions in human hand
  movements}.{\BBCQ}
\newblock
\APACjournalVolNumPages{Biological Cybernetics}{51}{5}{347--356}.
\PrintBackRefs{\CurrentBib}

\bibitem [\protect \citeauthoryear {%
Hastings%
}{%
Hastings%
}{%
{\protect \APACyear {1970}}%
}]{%
Hastings1970}
\APACinsertmetastar {%
Hastings1970}%
\begin{APACrefauthors}%
Hastings, W\BPBI K.%
\end{APACrefauthors}%
\unskip\
\newblock
\APACrefYearMonthDay{1970}{}{}.
\newblock
{\BBOQ}\APACrefatitle {Monte {C}arlo sampling methods using {M}arkov chains and
  their applications} {Monte {C}arlo sampling methods using {M}arkov chains and
  their applications}.{\BBCQ}
\newblock
\APACjournalVolNumPages{Biometrika}{57}{1}{97-109}.
\PrintBackRefs{\CurrentBib}

\bibitem [\protect \citeauthoryear {%
Heister%
\ \protect \BOthers {.}}{%
Heister%
\ \protect \BOthers {.}}{%
{\protect \APACyear {2011}}%
}]{%
Heister2011}
\APACinsertmetastar {%
Heister2011}%
\begin{APACrefauthors}%
Heister, J.%
, W{\"u}rzner, K\BHBI M.%
, Bubenzer, J.%
, Pohl, E.%
, Hanneforth, T.%
, Geyken, A.%
\BCBL {}\ \BBA {} Kliegl, R.%
\end{APACrefauthors}%
\unskip\
\newblock
\APACrefYearMonthDay{2011}{}{}.
\newblock
{\BBOQ}\APACrefatitle {dlex{DB}--eine lexikalische {D}atenbank f{\"u}r die
  psychologische und linguistische {F}orschung} {dlex{DB}--eine lexikalische
  {D}atenbank f{\"u}r die psychologische und linguistische {F}orschung}.{\BBCQ}
\newblock
\APACjournalVolNumPages{Psychologische Rundschau}{62}{1}{10--20}.
\PrintBackRefs{\CurrentBib}

\bibitem [\protect \citeauthoryear {%
Holmes%
}{%
Holmes%
}{%
{\protect \APACyear {2015}}%
}]{%
Holmes2015}
\APACinsertmetastar {%
Holmes2015}%
\begin{APACrefauthors}%
Holmes, W\BPBI R.%
\end{APACrefauthors}%
\unskip\
\newblock
\APACrefYearMonthDay{2015}{}{}.
\newblock
{\BBOQ}\APACrefatitle {A practical guide to the Probability Density
  Approximation (PDA) with improved implementation and error characterization}
  {A practical guide to the probability density approximation (pda) with
  improved implementation and error characterization}.{\BBCQ}
\newblock
\APACjournalVolNumPages{Journal of Mathematical Psychology}{68-69}{}{13--24}.
\PrintBackRefs{\CurrentBib}

\bibitem [\protect \citeauthoryear {%
Kliegl%
, Grabner%
, Rolfs%
\BCBL {}\ \BBA {} Engbert%
}{%
Kliegl%
\ \protect \BOthers {.}}{%
{\protect \APACyear {2004}}%
}]{%
Kliegl2004}
\APACinsertmetastar {%
Kliegl2004}%
\begin{APACrefauthors}%
Kliegl, R.%
, Grabner, E.%
, Rolfs, M.%
\BCBL {}\ \BBA {} Engbert, R.%
\end{APACrefauthors}%
\unskip\
\newblock
\APACrefYearMonthDay{2004}{}{}.
\newblock
{\BBOQ}\APACrefatitle {Length, frequency, and predictability effects of words
  on eye movements in reading} {Length, frequency, and predictability effects
  of words on eye movements in reading}.{\BBCQ}
\newblock
\APACjournalVolNumPages{European Journal of Cognitive
  Psychology}{16}{1-2}{262--284}.
\PrintBackRefs{\CurrentBib}

\bibitem [\protect \citeauthoryear {%
Kr{\"u}gel%
\ \BBA {} Engbert%
}{%
Kr{\"u}gel%
\ \BBA {} Engbert%
}{%
{\protect \APACyear {2010}}%
}]{%
Kruegel2010}
\APACinsertmetastar {%
Kruegel2010}%
\begin{APACrefauthors}%
Kr{\"u}gel, A.%
\BCBT {}\ \BBA {} Engbert, R.%
\end{APACrefauthors}%
\unskip\
\newblock
\APACrefYearMonthDay{2010}{}{}.
\newblock
{\BBOQ}\APACrefatitle {The launch-site effect for skipped words during reading}
  {The launch-site effect for skipped words during reading}.{\BBCQ}
\newblock
\APACjournalVolNumPages{Vision Research}{50}{}{1532--1539}.
\PrintBackRefs{\CurrentBib}

\bibitem [\protect \citeauthoryear {%
Kr{\"u}gel%
\ \BBA {} Engbert%
}{%
Kr{\"u}gel%
\ \BBA {} Engbert%
}{%
{\protect \APACyear {2014}}%
}]{%
Kruegel2014}
\APACinsertmetastar {%
Kruegel2014}%
\begin{APACrefauthors}%
Kr{\"u}gel, A.%
\BCBT {}\ \BBA {} Engbert, R.%
\end{APACrefauthors}%
\unskip\
\newblock
\APACrefYearMonthDay{2014}{}{}.
\newblock
{\BBOQ}\APACrefatitle {A model of saccadic landing positions in reading under
  the influence of sensory noise} {A model of saccadic landing positions in
  reading under the influence of sensory noise}.{\BBCQ}
\newblock
\APACjournalVolNumPages{Visual Cognition}{22}{3-4}{334--353}.
\PrintBackRefs{\CurrentBib}

\bibitem [\protect \citeauthoryear {%
Laloy%
\ \BBA {} Vrugt%
}{%
Laloy%
\ \BBA {} Vrugt%
}{%
{\protect \APACyear {2012}}%
}]{%
Laloy2012}
\APACinsertmetastar {%
Laloy2012}%
\begin{APACrefauthors}%
Laloy, E.%
\BCBT {}\ \BBA {} Vrugt, J\BPBI A.%
\end{APACrefauthors}%
\unskip\
\newblock
\APACrefYearMonthDay{2012}{}{}.
\newblock
{\BBOQ}\APACrefatitle {High-dimensional posterior exploration of hydrologic
  models using multiple-try DREAM (ZS) and high-performance computing}
  {High-dimensional posterior exploration of hydrologic models using
  multiple-try dream (zs) and high-performance computing}.{\BBCQ}
\newblock
\APACjournalVolNumPages{Water Resources Research}{48}{1}{}.
\PrintBackRefs{\CurrentBib}

\bibitem [\protect \citeauthoryear {%
Law%
, Stuart%
\BCBL {}\ \BBA {} Zygalakis%
}{%
Law%
\ \protect \BOthers {.}}{%
{\protect \APACyear {2015}}%
}]{%
Law2015}
\APACinsertmetastar {%
Law2015}%
\begin{APACrefauthors}%
Law, K.%
, Stuart, A.%
\BCBL {}\ \BBA {} Zygalakis, K.%
\end{APACrefauthors}%
\unskip\
\newblock
\APACrefYear{2015}.
\newblock
\APACrefbtitle {Data {A}ssimilation} {Data {A}ssimilation}.
\newblock
\APACaddressPublisher{}{Springer}.
\PrintBackRefs{\CurrentBib}

\bibitem [\protect \citeauthoryear {%
Marin%
\ \BBA {} Robert%
}{%
Marin%
\ \BBA {} Robert%
}{%
{\protect \APACyear {2007}}%
}]{%
Marin2007}
\APACinsertmetastar {%
Marin2007}%
\begin{APACrefauthors}%
Marin, J\BHBI M.%
\BCBT {}\ \BBA {} Robert, C.%
\end{APACrefauthors}%
\unskip\
\newblock
\APACrefYear{2007}.
\newblock
\APACrefbtitle {Bayesian {C}ore: {A} {P}ractical {A}pproach to {C}omputational
  {B}ayesian {S}tatistics} {Bayesian {C}ore: {A} {P}ractical {A}pproach to
  {C}omputational {B}ayesian {S}tatistics}.
\newblock
\APACaddressPublisher{}{Springer Science \& Business Media}.
\PrintBackRefs{\CurrentBib}

\bibitem [\protect \citeauthoryear {%
Matin%
}{%
Matin%
}{%
{\protect \APACyear {1974}}%
}]{%
Matin1974}
\APACinsertmetastar {%
Matin1974}%
\begin{APACrefauthors}%
Matin, E.%
\end{APACrefauthors}%
\unskip\
\newblock
\APACrefYearMonthDay{1974}{}{}.
\newblock
{\BBOQ}\APACrefatitle {Saccadic suppression: A review and an analysis.}
  {Saccadic suppression: A review and an analysis.}{\BBCQ}
\newblock
\APACjournalVolNumPages{Psychological Bulletin}{81}{12}{899--917}.
\PrintBackRefs{\CurrentBib}

\bibitem [\protect \citeauthoryear {%
McConkie%
, Kerr%
, Reddix%
\BCBL {}\ \BBA {} Zola%
}{%
McConkie%
\ \protect \BOthers {.}}{%
{\protect \APACyear {1988}}%
}]{%
McConkie1988}
\APACinsertmetastar {%
McConkie1988}%
\begin{APACrefauthors}%
McConkie, G\BPBI W.%
, Kerr, P\BPBI W.%
, Reddix, M\BPBI D.%
\BCBL {}\ \BBA {} Zola, D.%
\end{APACrefauthors}%
\unskip\
\newblock
\APACrefYearMonthDay{1988}{}{}.
\newblock
{\BBOQ}\APACrefatitle {Eye movement control during reading: I. {T}he location
  of initial eye fixations on words} {Eye movement control during reading: I.
  {T}he location of initial eye fixations on words}.{\BBCQ}
\newblock
\APACjournalVolNumPages{Vision Research}{28}{10}{1107--1118}.
\PrintBackRefs{\CurrentBib}

\bibitem [\protect \citeauthoryear {%
Myung%
}{%
Myung%
}{%
{\protect \APACyear {2003}}%
}]{%
Myung2003}
\APACinsertmetastar {%
Myung2003}%
\begin{APACrefauthors}%
Myung, I\BPBI J.%
\end{APACrefauthors}%
\unskip\
\newblock
\APACrefYearMonthDay{2003}{}{}.
\newblock
{\BBOQ}\APACrefatitle {Tutorial on maximum likelihood estimation} {Tutorial on
  maximum likelihood estimation}.{\BBCQ}
\newblock
\APACjournalVolNumPages{Journal of Mathematical Psychology}{47}{1}{90--100}.
\PrintBackRefs{\CurrentBib}

\bibitem [\protect \citeauthoryear {%
Nuthmann%
\ \BBA {} Engbert%
}{%
Nuthmann%
\ \BBA {} Engbert%
}{%
{\protect \APACyear {2009}}%
}]{%
Nuthmann2009}
\APACinsertmetastar {%
Nuthmann2009}%
\begin{APACrefauthors}%
Nuthmann, A.%
\BCBT {}\ \BBA {} Engbert, R.%
\end{APACrefauthors}%
\unskip\
\newblock
\APACrefYearMonthDay{2009}{}{}.
\newblock
{\BBOQ}\APACrefatitle {Mindless reading revisited: An analysis based on the
  SWIFT model of eye-movement control} {Mindless reading revisited: An analysis
  based on the swift model of eye-movement control}.{\BBCQ}
\newblock
\APACjournalVolNumPages{Vision Research}{49}{3}{322--336}.
\PrintBackRefs{\CurrentBib}

\bibitem [\protect \citeauthoryear {%
Nuthmann%
, Engbert%
\BCBL {}\ \BBA {} Kliegl%
}{%
Nuthmann%
\ \protect \BOthers {.}}{%
{\protect \APACyear {2005}}%
}]{%
Nuthmann2005}
\APACinsertmetastar {%
Nuthmann2005}%
\begin{APACrefauthors}%
Nuthmann, A.%
, Engbert, R.%
\BCBL {}\ \BBA {} Kliegl, R.%
\end{APACrefauthors}%
\unskip\
\newblock
\APACrefYearMonthDay{2005}{}{}.
\newblock
{\BBOQ}\APACrefatitle {Mislocated fixations during reading and the inverted
  optimal viewing position effect} {Mislocated fixations during reading and the
  inverted optimal viewing position effect}.{\BBCQ}
\newblock
\APACjournalVolNumPages{Vision Research}{45}{17}{2201--2217}.
\PrintBackRefs{\CurrentBib}

\bibitem [\protect \citeauthoryear {%
Palestro%
, Sederberg%
, Osth%
, Van~Zandt%
\BCBL {}\ \BBA {} Turner%
}{%
Palestro%
\ \protect \BOthers {.}}{%
{\protect \APACyear {2018}}%
}]{%
Palestro2018}
\APACinsertmetastar {%
Palestro2018}%
\begin{APACrefauthors}%
Palestro, J\BPBI J.%
, Sederberg, P\BPBI B.%
, Osth, A\BPBI F.%
, Van~Zandt, T.%
\BCBL {}\ \BBA {} Turner, B\BPBI M.%
\end{APACrefauthors}%
\unskip\
\newblock
\APACrefYear{2018}.
\newblock
\APACrefbtitle {Likelihood-free methods for cognitive science} {Likelihood-free
  methods for cognitive science}.
\newblock
\APACaddressPublisher{}{Springer}.
\PrintBackRefs{\CurrentBib}

\bibitem [\protect \citeauthoryear {%
Rayner%
}{%
Rayner%
}{%
{\protect \APACyear {1975}}%
}]{%
Rayner1975}
\APACinsertmetastar {%
Rayner1975}%
\begin{APACrefauthors}%
Rayner, K.%
\end{APACrefauthors}%
\unskip\
\newblock
\APACrefYearMonthDay{1975}{}{}.
\newblock
{\BBOQ}\APACrefatitle {The perceptual span and peripheral cues in reading} {The
  perceptual span and peripheral cues in reading}.{\BBCQ}
\newblock
\APACjournalVolNumPages{Cognitive Psychology}{7}{1}{65--81}.
\PrintBackRefs{\CurrentBib}

\bibitem [\protect \citeauthoryear {%
Rayner%
}{%
Rayner%
}{%
{\protect \APACyear {1998}}%
}]{%
Rayner1998}
\APACinsertmetastar {%
Rayner1998}%
\begin{APACrefauthors}%
Rayner, K.%
\end{APACrefauthors}%
\unskip\
\newblock
\APACrefYearMonthDay{1998}{}{}.
\newblock
{\BBOQ}\APACrefatitle {Eye movements in reading and information processing: 20
  years of research.} {Eye movements in reading and information processing: 20
  years of research.}{\BBCQ}
\newblock
\APACjournalVolNumPages{Psychological Bulletin}{124}{3}{372--422}.
\PrintBackRefs{\CurrentBib}

\bibitem [\protect \citeauthoryear {%
Rayner%
\ \BBA {} Reichle%
}{%
Rayner%
\ \BBA {} Reichle%
}{%
{\protect \APACyear {2010}}%
}]{%
Rayner2010}
\APACinsertmetastar {%
Rayner2010}%
\begin{APACrefauthors}%
Rayner, K.%
\BCBT {}\ \BBA {} Reichle, E\BPBI D.%
\end{APACrefauthors}%
\unskip\
\newblock
\APACrefYearMonthDay{2010}{}{}.
\newblock
{\BBOQ}\APACrefatitle {Models of the reading process} {Models of the reading
  process}.{\BBCQ}
\newblock
\APACjournalVolNumPages{Wiley Interdisciplinary Reviews: Cognitive
  Science}{1}{6}{787--799}.
\PrintBackRefs{\CurrentBib}

\bibitem [\protect \citeauthoryear {%
Rayner%
, Well%
\BCBL {}\ \BBA {} Pollatsek%
}{%
Rayner%
\ \protect \BOthers {.}}{%
{\protect \APACyear {1980}}%
}]{%
Rayner1980}
\APACinsertmetastar {%
Rayner1980}%
\begin{APACrefauthors}%
Rayner, K.%
, Well, A\BPBI D.%
\BCBL {}\ \BBA {} Pollatsek, A.%
\end{APACrefauthors}%
\unskip\
\newblock
\APACrefYearMonthDay{1980}{}{}.
\newblock
{\BBOQ}\APACrefatitle {Asymmetry of the effective visual field in reading}
  {Asymmetry of the effective visual field in reading}.{\BBCQ}
\newblock
\APACjournalVolNumPages{Perception \& Psychophysics}{27}{6}{537--544}.
\PrintBackRefs{\CurrentBib}

\bibitem [\protect \citeauthoryear {%
Reich%
\ \BBA {} Cotter%
}{%
Reich%
\ \BBA {} Cotter%
}{%
{\protect \APACyear {2015}}%
}]{%
Reich2015}
\APACinsertmetastar {%
Reich2015}%
\begin{APACrefauthors}%
Reich, S.%
\BCBT {}\ \BBA {} Cotter, C.%
\end{APACrefauthors}%
\unskip\
\newblock
\APACrefYear{2015}.
\newblock
\APACrefbtitle {Probabilistic {F}orecasting and {B}ayesian {D}ata
  {A}ssimilation} {Probabilistic {F}orecasting and {B}ayesian {D}ata
  {A}ssimilation}.
\newblock
\APACaddressPublisher{}{Cambridge University Press}.
\PrintBackRefs{\CurrentBib}

\bibitem [\protect \citeauthoryear {%
Reichle%
, Pollatsek%
, Fisher%
\BCBL {}\ \BBA {} Rayner%
}{%
Reichle%
\ \protect \BOthers {.}}{%
{\protect \APACyear {1998}}%
}]{%
Reichle1998}
\APACinsertmetastar {%
Reichle1998}%
\begin{APACrefauthors}%
Reichle, E\BPBI D.%
, Pollatsek, A.%
, Fisher, D\BPBI L.%
\BCBL {}\ \BBA {} Rayner, K.%
\end{APACrefauthors}%
\unskip\
\newblock
\APACrefYearMonthDay{1998}{}{}.
\newblock
{\BBOQ}\APACrefatitle {Toward a model of eye movement control in reading.}
  {Toward a model of eye movement control in reading.}{\BBCQ}
\newblock
\APACjournalVolNumPages{Psychological Review}{105}{1}{125--157}.
\PrintBackRefs{\CurrentBib}

\bibitem [\protect \citeauthoryear {%
Reichle%
, Rayner%
\BCBL {}\ \BBA {} Pollatsek%
}{%
Reichle%
\ \protect \BOthers {.}}{%
{\protect \APACyear {2003}}%
}]{%
Reichle2003}
\APACinsertmetastar {%
Reichle2003}%
\begin{APACrefauthors}%
Reichle, E\BPBI D.%
, Rayner, K.%
\BCBL {}\ \BBA {} Pollatsek, A.%
\end{APACrefauthors}%
\unskip\
\newblock
\APACrefYearMonthDay{2003}{}{}.
\newblock
{\BBOQ}\APACrefatitle {The {E}-{Z} {R}eader model of eye-movement control in
  reading: Comparisons to other models} {The {E}-{Z} {R}eader model of
  eye-movement control in reading: Comparisons to other models}.{\BBCQ}
\newblock
\APACjournalVolNumPages{Behavioral and Brain Sciences}{26}{4}{445--476}.
\PrintBackRefs{\CurrentBib}

\bibitem [\protect \citeauthoryear {%
Reichle%
, Warren%
\BCBL {}\ \BBA {} McConnell%
}{%
Reichle%
\ \protect \BOthers {.}}{%
{\protect \APACyear {2009}}%
}]{%
Reichle2009}
\APACinsertmetastar {%
Reichle2009}%
\begin{APACrefauthors}%
Reichle, E\BPBI D.%
, Warren, T.%
\BCBL {}\ \BBA {} McConnell, K.%
\end{APACrefauthors}%
\unskip\
\newblock
\APACrefYearMonthDay{2009}{}{}.
\newblock
{\BBOQ}\APACrefatitle {Using {E-Z} {R}eader to model the effects of higher
  level language processing on eye movements during reading} {Using {E-Z}
  {R}eader to model the effects of higher level language processing on eye
  movements during reading}.{\BBCQ}
\newblock
\APACjournalVolNumPages{Psychonomic Bulletin \& Review}{16}{1}{1--21}.
\PrintBackRefs{\CurrentBib}

\bibitem [\protect \citeauthoryear {%
Risse%
}{%
Risse%
}{%
{\protect \APACyear {2014}}%
}]{%
Risse2014}
\APACinsertmetastar {%
Risse2014}%
\begin{APACrefauthors}%
Risse, S.%
\end{APACrefauthors}%
\unskip\
\newblock
\APACrefYearMonthDay{2014}{}{}.
\newblock
{\BBOQ}\APACrefatitle {Effects of visual span on reading speed and parafoveal
  processing in eye movements during sentence reading} {Effects of visual span
  on reading speed and parafoveal processing in eye movements during sentence
  reading}.{\BBCQ}
\newblock
\APACjournalVolNumPages{Journal of Vision}{14}{8}{11--11}.
\PrintBackRefs{\CurrentBib}

\bibitem [\protect \citeauthoryear {%
Risse%
\ \BBA {} Seelig%
}{%
Risse%
\ \BBA {} Seelig%
}{%
{\protect \APACyear {2019}}%
}]{%
RisseSeeligQJEP}
\APACinsertmetastar {%
RisseSeeligQJEP}%
\begin{APACrefauthors}%
Risse, S.%
\BCBT {}\ \BBA {} Seelig, S.%
\end{APACrefauthors}%
\unskip\
\newblock
\APACrefYearMonthDay{2019}{}{}.
\newblock
{\BBOQ}\APACrefatitle {Stable preview difficulty effects in reading with an
  improved variant of the boundary paradigm} {Stable preview difficulty effects
  in reading with an improved variant of the boundary paradigm}.{\BBCQ}
\newblock
\APACjournalVolNumPages{Quarterly Journal of Experimental Psychology}{}{}{}.
\PrintBackRefs{\CurrentBib}

\bibitem [\protect \citeauthoryear {%
Robert%
\ \BBA {} Casella%
}{%
Robert%
\ \BBA {} Casella%
}{%
{\protect \APACyear {2013}}%
}]{%
Robert2013}
\APACinsertmetastar {%
Robert2013}%
\begin{APACrefauthors}%
Robert, C.%
\BCBT {}\ \BBA {} Casella, G.%
\end{APACrefauthors}%
\unskip\
\newblock
\APACrefYear{2013}.
\newblock
\APACrefbtitle {Monte {C}arlo {S}tatistical {M}ethods} {Monte {C}arlo
  {S}tatistical {M}ethods}.
\newblock
\APACaddressPublisher{}{Springer Science \& Business Media}.
\PrintBackRefs{\CurrentBib}

\bibitem [\protect \citeauthoryear {%
Roberts%
, Gelman%
\BCBL {}\ \BBA {} Gilks%
}{%
Roberts%
\ \protect \BOthers {.}}{%
{\protect \APACyear {1997}}%
}]{%
Roberts1997}
\APACinsertmetastar {%
Roberts1997}%
\begin{APACrefauthors}%
Roberts, G\BPBI O.%
, Gelman, A.%
\BCBL {}\ \BBA {} Gilks, W\BPBI R.%
\end{APACrefauthors}%
\unskip\
\newblock
\APACrefYearMonthDay{1997}{}{}.
\newblock
{\BBOQ}\APACrefatitle {Weak convergence and optimal scaling of random walk
  Metropolis algorithms} {Weak convergence and optimal scaling of random walk
  metropolis algorithms}.{\BBCQ}
\newblock
\APACjournalVolNumPages{The Annals of Applied Probability}{7}{1}{110--120}.
\PrintBackRefs{\CurrentBib}

\bibitem [\protect \citeauthoryear {%
Schad%
\ \BBA {} Engbert%
}{%
Schad%
\ \BBA {} Engbert%
}{%
{\protect \APACyear {2012}}%
}]{%
Schad2012}
\APACinsertmetastar {%
Schad2012}%
\begin{APACrefauthors}%
Schad, D\BPBI J.%
\BCBT {}\ \BBA {} Engbert, R.%
\end{APACrefauthors}%
\unskip\
\newblock
\APACrefYearMonthDay{2012}{}{}.
\newblock
{\BBOQ}\APACrefatitle {The zoom lens of attention: Simulating shuffled versus
  normal text reading using the {SWIFT} model} {The zoom lens of attention:
  Simulating shuffled versus normal text reading using the {SWIFT}
  model}.{\BBCQ}
\newblock
\APACjournalVolNumPages{Visual Cognition}{20}{4-5}{391--421}.
\PrintBackRefs{\CurrentBib}

\bibitem [\protect \citeauthoryear {%
Sch{\"u}tt%
\ \protect \BOthers {.}}{%
Sch{\"u}tt%
\ \protect \BOthers {.}}{%
{\protect \APACyear {2017}}%
}]{%
Schuett2017}
\APACinsertmetastar {%
Schuett2017}%
\begin{APACrefauthors}%
Sch{\"u}tt, H\BPBI H.%
, Rothkegel, L\BPBI O.%
, Trukenbrod, H\BPBI A.%
, Reich, S.%
, Wichmann, F\BPBI A.%
\BCBL {}\ \BBA {} Engbert, R.%
\end{APACrefauthors}%
\unskip\
\newblock
\APACrefYearMonthDay{2017}{}{}.
\newblock
{\BBOQ}\APACrefatitle {Likelihood-based parameter estimation and comparison of
  dynamical cognitive models.} {Likelihood-based parameter estimation and
  comparison of dynamical cognitive models.}{\BBCQ}
\newblock
\APACjournalVolNumPages{Psychological Review}{124}{4}{505--524}.
\PrintBackRefs{\CurrentBib}

\bibitem [\protect \citeauthoryear {%
Scott%
}{%
Scott%
}{%
{\protect \APACyear {2015}}%
}]{%
Scott2015}
\APACinsertmetastar {%
Scott2015}%
\begin{APACrefauthors}%
Scott, D\BPBI W.%
\end{APACrefauthors}%
\unskip\
\newblock
\APACrefYear{2015}.
\newblock
\APACrefbtitle {Multivariate {D}ensity {E}stimation: {T}heory, {P}ractice, and
  {V}isualization} {Multivariate {D}ensity {E}stimation: {T}heory, {P}ractice,
  and {V}isualization}.
\newblock
\APACaddressPublisher{}{John Wiley \& Sons}.
\PrintBackRefs{\CurrentBib}

\bibitem [\protect \citeauthoryear {%
Sisson%
\ \BBA {} Fan%
}{%
Sisson%
\ \BBA {} Fan%
}{%
{\protect \APACyear {2011}}%
}]{%
Sisson2011}
\APACinsertmetastar {%
Sisson2011}%
\begin{APACrefauthors}%
Sisson, S\BPBI A.%
\BCBT {}\ \BBA {} Fan, Y.%
\end{APACrefauthors}%
\unskip\
\newblock
\APACrefYearMonthDay{2011}{}{}.
\newblock
{\BBOQ}\APACrefatitle {Likelihood-free {M}arkov chain {M}onte {C}arlo}
  {Likelihood-free {M}arkov chain {M}onte {C}arlo}.{\BBCQ}
\newblock
\BIn{} (\BPGS\ 313--335).
\newblock
\APACaddressPublisher{}{Chapman \& Hall/CRC, New York}.
\PrintBackRefs{\CurrentBib}

\bibitem [\protect \citeauthoryear {%
ter Braak%
}{%
ter Braak%
}{%
{\protect \APACyear {2006}}%
}]{%
Ter2006}
\APACinsertmetastar {%
Ter2006}%
\begin{APACrefauthors}%
ter Braak, C\BPBI J.%
\end{APACrefauthors}%
\unskip\
\newblock
\APACrefYearMonthDay{2006}{}{}.
\newblock
{\BBOQ}\APACrefatitle {A {M}arkov Chain {M}onte {C}arlo version of the genetic
  algorithm Differential Evolution: {E}asy {B}ayesian computing for real
  parameter spaces} {A {M}arkov chain {M}onte {C}arlo version of the genetic
  algorithm differential evolution: {E}asy {B}ayesian computing for real
  parameter spaces}.{\BBCQ}
\newblock
\APACjournalVolNumPages{Statistics and Computing}{16}{3}{239--249}.
\PrintBackRefs{\CurrentBib}

\bibitem [\protect \citeauthoryear {%
ter Braak%
\ \BBA {} Vrugt%
}{%
ter Braak%
\ \BBA {} Vrugt%
}{%
{\protect \APACyear {2008}}%
}]{%
Ter2008}
\APACinsertmetastar {%
Ter2008}%
\begin{APACrefauthors}%
ter Braak, C\BPBI J.%
\BCBT {}\ \BBA {} Vrugt, J\BPBI A.%
\end{APACrefauthors}%
\unskip\
\newblock
\APACrefYearMonthDay{2008}{}{}.
\newblock
{\BBOQ}\APACrefatitle {Differential evolution {M}arkov chain with snooker
  updater and fewer chains} {Differential evolution {M}arkov chain with snooker
  updater and fewer chains}.{\BBCQ}
\newblock
\APACjournalVolNumPages{Statistics and Computing}{18}{4}{435--446}.
\PrintBackRefs{\CurrentBib}

\bibitem [\protect \citeauthoryear {%
Turner%
\ \BBA {} Sederberg%
}{%
Turner%
\ \BBA {} Sederberg%
}{%
{\protect \APACyear {2014}}%
}]{%
Turner2014}
\APACinsertmetastar {%
Turner2014}%
\begin{APACrefauthors}%
Turner, B\BPBI M.%
\BCBT {}\ \BBA {} Sederberg, P\BPBI B.%
\end{APACrefauthors}%
\unskip\
\newblock
\APACrefYearMonthDay{2014}{}{}.
\newblock
{\BBOQ}\APACrefatitle {A generalized, likelihood-free method for posterior
  estimation} {A generalized, likelihood-free method for posterior
  estimation}.{\BBCQ}
\newblock
\APACjournalVolNumPages{Psychonomic bulletin \& review}{21}{2}{227--250}.
\PrintBackRefs{\CurrentBib}

\bibitem [\protect \citeauthoryear {%
Van~Gelder%
}{%
Van~Gelder%
}{%
{\protect \APACyear {1998}}%
}]{%
vanGelder1998}
\APACinsertmetastar {%
vanGelder1998}%
\begin{APACrefauthors}%
Van~Gelder, T.%
\end{APACrefauthors}%
\unskip\
\newblock
\APACrefYearMonthDay{1998}{}{}.
\newblock
{\BBOQ}\APACrefatitle {The dynamical hypothesis in cognitive science} {The
  dynamical hypothesis in cognitive science}.{\BBCQ}
\newblock
\APACjournalVolNumPages{Behavioral and Brain Sciences}{21}{5}{615--628}.
\PrintBackRefs{\CurrentBib}

\bibitem [\protect \citeauthoryear {%
Van~Kampen%
}{%
Van~Kampen%
}{%
{\protect \APACyear {1992}}%
}]{%
vanKampen1992}
\APACinsertmetastar {%
vanKampen1992}%
\begin{APACrefauthors}%
Van~Kampen, N\BPBI G.%
\end{APACrefauthors}%
\unskip\
\newblock
\APACrefYear{1992}.
\newblock
\APACrefbtitle {Stochastic {P}rocesses in {P}hysics and {C}hemistry}
  {Stochastic {P}rocesses in {P}hysics and {C}hemistry}\ (\BVOL~1).
\newblock
\APACaddressPublisher{}{Elsevier: North-Holland}.
\PrintBackRefs{\CurrentBib}

\bibitem [\protect \citeauthoryear {%
Vasilev%
\ \BBA {} Angele%
}{%
Vasilev%
\ \BBA {} Angele%
}{%
{\protect \APACyear {2017}}%
}]{%
VasilevAngele2017}
\APACinsertmetastar {%
VasilevAngele2017}%
\begin{APACrefauthors}%
Vasilev, M\BPBI R.%
\BCBT {}\ \BBA {} Angele, B.%
\end{APACrefauthors}%
\unskip\
\newblock
\APACrefYearMonthDay{2017}{}{}.
\newblock
{\BBOQ}\APACrefatitle {Parafoveal preview effects from word N+ 1 and word N+ 2
  during reading: A critical review and {B}ayesian meta-analysis} {Parafoveal
  preview effects from word n+ 1 and word n+ 2 during reading: A critical
  review and {B}ayesian meta-analysis}.{\BBCQ}
\newblock
\APACjournalVolNumPages{Psychonomic Bulletin \& Review}{24}{3}{666--689}.
\PrintBackRefs{\CurrentBib}

\bibitem [\protect \citeauthoryear {%
Vihola%
}{%
Vihola%
}{%
{\protect \APACyear {2012}}%
}]{%
Vihola2012}
\APACinsertmetastar {%
Vihola2012}%
\begin{APACrefauthors}%
Vihola, M.%
\end{APACrefauthors}%
\unskip\
\newblock
\APACrefYearMonthDay{2012}{}{}.
\newblock
{\BBOQ}\APACrefatitle {Robust adaptive Metropolis algorithm with coerced
  acceptance rate} {Robust adaptive metropolis algorithm with coerced
  acceptance rate}.{\BBCQ}
\newblock
\APACjournalVolNumPages{Statistics and Computing}{22}{5}{997--1008}.
\PrintBackRefs{\CurrentBib}

\bibitem [\protect \citeauthoryear {%
Vitu%
, McConkie%
, Kerr%
\BCBL {}\ \BBA {} O'Regan%
}{%
Vitu%
\ \protect \BOthers {.}}{%
{\protect \APACyear {2001}}%
}]{%
Vitu2001}
\APACinsertmetastar {%
Vitu2001}%
\begin{APACrefauthors}%
Vitu, F.%
, McConkie, G\BPBI W.%
, Kerr, P.%
\BCBL {}\ \BBA {} O'Regan, J\BPBI K.%
\end{APACrefauthors}%
\unskip\
\newblock
\APACrefYearMonthDay{2001}{}{}.
\newblock
{\BBOQ}\APACrefatitle {Fixation location effects on fixation durations during
  reading: An inverted optimal viewing position effect} {Fixation location
  effects on fixation durations during reading: An inverted optimal viewing
  position effect}.{\BBCQ}
\newblock
\APACjournalVolNumPages{Vision Research}{41}{25-26}{3513--3533}.
\PrintBackRefs{\CurrentBib}

\end{thebibliography}

\begin{appendix}
\section*{Appendix: Experimental data and sentence material} 
\label{appx_expdata}
All eye-tracking data used in our simulation studies originate from \cite{RisseSeeligQJEP}, who collected data for an experiment that was a version of the $n+1$ boundary paradigm \citep{Rayner1975} to investigate effects of parafoveal word difficulty on fixation durations and distinguish them from preview benefit effects (see \citeauthor{VasilevAngele2017}, \citeyear{VasilevAngele2017}, for a comprehensive review). \re{Their data is available online at \href{https://osf.io/kz483/}{10.17605/OSF.IO/KZ483}.}

In the experiment, 34 participants, mostly students of psychology at the University of Potsdam, read 114 single sentences presented on a computer screen while their eyes were being tracked. The simple structured German sentences consisted of six to 12 words with an average length of 9 words. Every sentence contained a gaze contingent invisible boundary before a specific target word. Before the eyes crossed the boundary, the preview of the target word could either be of low, high or medium frequency (i.e. high, low or medium difficulty respectively). During the saccade in which the boundary was crossed, the target word was always exchanged with the medium frequency word. Word frequencies were taken from the dlexDB database~\citep{Heister2011} based on \textit{The DWDS corpus: A reference corpus for the German language of the 20th century}~\citep{Geyken2007}. 

\textit{Data treatment and preprocessing.} The data were collected using an Eyelink II System (\textit{SR Research, Osgoode/Ontario, Canada}) with a temporal resolution of 1,000~Hz. Since spatial resolution was preprocessed to letter accuracy. Within-letter position was randomized by added small random numbers to avoid artifacts from discretization. Basically, the data used here were treated by the same preprocessing as reported in the statistical analysis of the experiment. Additionally, fixation durations smaller than 25~ms were discarded (550 fixations in 338 trials). Trials that included fixation durations larger than 1,000~ms were discarded (45). Trials consisting of less than three fixations were also removed from the data-set. Additionally, re-readings signaled by regressions starting from the second last or last word of the sentence and all subsequent fixations were discarded (5,773 fixations). After preprocessing, 30,639 fixations from 3,422 trials were included in the data-set for estimation. 
\re{The implementations of the model, the estimation algorithm, and scripts for analyses and plots, along with the corpus data and fixation sequences are available at \href{https://osf.io/xdkwq/}{10.17605/OSF.IO/XDKWQ}.}
\end{appendix}

\end{document}